\documentclass[conference]{IEEEtran}
\IEEEoverridecommandlockouts
\usepackage[frozencache,cachedir=.]{minted}
\usepackage{cite}
\usepackage{amsmath,amssymb,amsfonts}
\usepackage{algorithmic}
\usepackage{graphicx}
\usepackage{textcomp}
\usepackage{xcolor}
\usepackage{pifont}
\newcommand{\revised}[1]{{\color{black}{#1}}}
\usepackage[utf8]{inputenc}
\usepackage{mathtools}
\usepackage{caption}
\usepackage{subcaption}
\usepackage{comment}
\usepackage{graphicx}
\usepackage{tikz}
\usepackage{url}
\usepackage{verbatim}
\usepackage{todonotes}
\usepackage{balance}
\usepackage{xcolor}
\usepackage{amsmath}
\usepackage{amsthm}
\usepackage{amssymb}
\usepackage{amssymb}
\usepackage{algorithm2e} 
\usepackage[listings]{tcolorbox}
\usepackage[numbers]{natbib}
\tcbuselibrary{all}
\usepackage{balance}
\newtheorem{definition}{Definition}
\def\BibTeX{{\rm B\kern-.05em{\sc i\kern-.025em b}\kern-.08em
    T\kern-.1667em\lower.7ex\hbox{E}\kern-.125emX}}
\newcommand{\uL}{tk}
\newcommand{\uP}{token}

\newcommand{\negl}{\ensuremath{\mathsf{negl}(\lambda)}\xspace}

\newcommand{\squishlist}{
 \begin{list}{$\bullet$}
  { \setlength{\itemsep}{0pt}
     \setlength{\parsep}{2pt}
     \setlength{\topsep}{3pt}
     \setlength{\partopsep}{0pt}
     \setlength{\leftmargin}{1em}
     \setlength{\labelwidth}{1em}
     \setlength{\labelsep}{0.5em} } }

\newcommand{\squishlisttwo}{
 \begin{list}{$\bullet$}
  { \setlength{\itemsep}{0pt}
    \setlength{\parsep}{0pt}
    \setlength{	opsep}{0pt}
    \setlength{\partopsep}{0pt}
    \setlength{\leftmargin}{2em}
    \setlength{\labelwidth}{1.5em}
    \setlength{\labelsep}{0.5em} } }

\newcommand{\squishend}{
  \end{list}  }

\newtcolorbox[auto counter,number format=\Roman ]{freqywm}[2][]{enhanced,colback=white,
fonttitle=\bfseries,coltitle=gray!25!black,
attach boxed title to top left=
{xshift=2mm,yshift=-3mm,yshifttext=-1mm},
boxed title style={colframe=gray!75!black,
colback=yellow!50!gray},
title=Algorithm \thetcbcounter: #2,#1}

\newtcolorbox[auto counter,number format=\Roman ]{securitygame}[2][]{enhanced,colback=white,
fonttitle=\bfseries,coltitle=gray!25!black,
attach boxed title to top center={yshift=-3mm},,
boxed title style={colframe=gray!75!black,
colback=yellow!50!gray},
title=Security Game \thetcbcounter: #2,#1}
\begin{document}
%\input{ICDE/revision_response}

%\pagebreak
\title{FreqyWM: Frequency Watermarking for the New Data Economy
}

\author{\IEEEauthorblockN{Devriş İşler\textsuperscript{1,2}, Elisa Cabana\textsuperscript{3}\textsuperscript{\textsection}, Alvaro Garcia-Recuero\textsuperscript{4}\textsuperscript{\textsection}, Georgia Koutrika\textsuperscript{5}, and Nikolaos Laoutaris\textsuperscript{1}}
\IEEEauthorblockA{\textit{\textsuperscript{1}IMDEA Networks Institute, \textsuperscript{2}UC3M, \textsuperscript{3}CUNEF University, \textsuperscript{4}FUNDITEC, \textsuperscript{5}Athena Research Center
}  \\
\{\small \textit{devris.isler, nikolaos.laoutaris\}@imdea.org, elisa.cabana@cunef.edu, algarcia@funditec.es, georgia@athenarc.gr} \\
\textit{\color{blue}*Please cite the ICDE'24 version of this paper.}
}
}
\maketitle
\begingroup\renewcommand\thefootnote{\textsection}
\footnotetext{Work done while the author was affiliated with IMDEA Networks Institute.}
\endgroup
\begin{abstract}
We present a novel technique for modulating the appearance frequency of a few tokens within a dataset for encoding an invisible watermark that can be used to protect ownership rights upon data. We develop optimal as well as fast heuristic algorithms for creating and verifying such watermarks. We also demonstrate the robustness of our technique against various attacks and derive analytical bounds for the false positive probability of erroneously ``detecting'' a watermark on a dataset that does not carry it. Our technique is applicable to both single dimensional and multidimensional datasets, is independent of token type, allows for a fine control of the introduced distortion, and can be used in a variety of use cases that involve buying and selling data in contemporary data marketplaces. 
\end{abstract}


\section{Introduction}\label{sec:intro}
Data-driven decision making powered by Machine Learning (ML) algorithms is changing how society and the economy work.
ML is driving up the demand for data in what has been called the fourth industrial revolution. 
To satisfy this demand, several data marketplaces (DMs), which are mediation platforms aiming to connect the two primary stakeholders of the data value chain, namely the data providers/sellers and the data buyers~\cite{azcoitia2022survey}, have appeared in the last few years. 

\noindent \textbf{The problems:}
Unfortunately, as with all digital assets, being able to copy/store/transmit datasets with close to zero cost makes creating illegal copies very easy. 
Even worse, unlike media content and software, the issue of ownership is less obvious when it comes to datasets. 
Any movie, song, e-book, or software can usually be attributed to a director, musician, author, or company, respectively, but this is hard to do for large datasets. 
\revised{These large datasets in data economy are traded in a wholesale manner that involves large numbers of tuples/rows.} 
Consider an anonymised mobility dataset logging the movement of people in a city. 
Such a dataset may have been produced by collecting GPS readings from the smartphones of individuals using a map application, or it may be deduced by analysing cell phone traces~\cite{10.1145/3419394.3423655} or Call Description Records (CDRs) maintained by mobile operators.
Deployment of advanced privacy enhancing technologies (PETs) such as multiparty computation \cite{mpcevans18}, (fully) homomorphic encryption \cite{fheGentry09}, functional encryption \cite{bonehfe11}, and trusted execution environments \cite{teewhatnot15} can protect data from leaking in the first place and allow (pre-agreed) computations on data without hampering the functioning of the data-driven economy, e.g., private set computation \cite{prism2021}, encrypted databases \cite{arxvldb2019}, secure computation \cite{teeAnciaux19}, secure data aggregation \cite{heda2022}, and verifiable databases \cite{veridb21}. 
However, most such approaches face serious scalability challenges that hamper their deployment in real-world use-cases. 
An alternative to deploying PETs solutions, is to rely on purely legal tools and terms and conditions to protect data ownership in the context of the new data economy~\cite{dataownershiplaw2022}. In fact, most DMs do exactly that -- trade plaintext versions of \revised{entire} data~\cite{azcoitia2022survey,dmrevisit22,dmassets20} assuming that the different parties will abide to pre-agreed terms and conditions.
With weak to nonexistent ownership guarantees by technical means, it is difficult to imagine that the data economy will ever flourish and reach its projected potential \cite{banterle2020data}. 
Indeed, any sold copy of a dataset can be `pirated' by a buyer-turned-seller that can then resell the same dataset in a DM thereby undercutting the rightful owner and rendering its investment useless. 

Watermarking is a well-known technique for protecting ownership upon copying and unauthorized distribution, initially proposed for protecting digital media \cite{asikuzzaman2017overview,begum2020digital} and software \cite{softwarewmtifs}. 
Watermarking techniques for datasets \cite{xmlwatermarking,agrawal2002watermarking,agrawalsigmod} and machine learning models \cite{wang2021riga} have been proposed recently. 
Watermarking generally consists of two algorithms: \textit{generation} (or embedding) and \textit{detection}. 
The generation allows an owner to embed an invisible (or visible) watermark into their data using a high entropy (watermarking) secret and produces a watermarked version of the data introducing tolerable distortion without degrading the data utility. 
During the detection algorithm, the owner proves its ownership on the suspected data (even if it is modified) using the same watermarking secret generated during the watermark generation. 
If the result of the detection is $1$ (or \textit{accept}), the owner can use it to prove their ownership on the (suspected) watermarked data.
A watermarking scheme is assumed to be secure against the guess attack (where an attacker tries to expose the watermarking secret) and robust against (un)intentional alterations/modifications (i.e., a watermark should be still detectable even under attacks such as ~\cite{RaniH22,agarwal2019survey,agrawal2003watermarking,agrawal2002watermarking,ji2021curse,quiring2018forgotten,cohen2018watermarking}).
 
\noindent \textbf{Limitations of existing watermarking techniques:} Watermarking techniques, depending on the nature of their application, may have very different objectives, e.g., numerical database watermarking controlling the distortion on mean and standard deviation \cite{agrawalsigmod}, reversible watermarking allowing owners to reconstruct the original data \cite{bigdb21}, watermarking text datasets preserving the meaning of a text \cite{dbwmli22} and/or the frequencies of the words \cite{textwm2021}, categorical watermarking preserving the (predefined) categories (e.g., gender) of a dataset \cite{categorywm21}.
All these solutions focus on a specific data type in a specific domain~\cite{kumar2020recent,RaniH22}. 
Another limitation of theirs relates to the level of control they offer to the user in terms of controlling the distortion introduced upon the original data due to the watermark. There are, for example, techniques that maintain the mean and the standard deviation of a numerical field~\cite{jony2021robust,agrawal2002watermarking,shehab2007watermarking} but, as we will show later, this can lead to arbitrary large distortion between the original and the watermarked data when considering the entire distribution of values that goes beyond the mean and the standard deviation.  

To address these limitations, \textit{we introduce a novel watermarking technique that can be implemented over a wide range of data types and structures }(with some constraints that will be explained later) while \textit{giving the data owner very precise control over the introduced distortion}. 
 
\noindent \textbf{A novel watermarking technique for data:} In this paper, we  present a novel \emph{Frequency Watermarking} technique, henceforth \textit{FreqyWM}, \footnote{Freqy pronounced as \textit{freaky}.} for hiding a secret within a dataset in a manner that makes the said secret indistinguishable from the data it protects. The main idea behind \textit{FreqyWM} is \textit{to slightly modify the appearance frequency of existing tokens} within a dataset in order \textit{to create a secret in the form of a complex relationship} between the frequencies of different tokens. By making this relationship complex enough, we can reduce the probability that it appeared by chance close to zero. Therefore, by revealing knowledge of such secret relationship, a party can claim ownership over a dataset because the only practical way of knowing such a secret is to have inserted it in the data in the first place. A token may be a word, a database record, a URL, or any repeating value within a structured or semi-structured commercial dataset. Our secret is created by first selecting a number of token pairs. Then, for each pair, we slightly modulate the frequency counts of its tokens in order to make their difference yield zero under modulo $N$ arithmetic. This can be easily done by adding or removing some instances of one, the other, or both tokens. By increasing the number of selected pairs we can make our watermark more resistant to attacks, as well as less likely to have appeared by chance. 

\textit{FreqyWM} can achieve several things. First and foremost, by revealing knowledge of the secret encoded by the watermark, a data seller can prove rightful ownership of a dataset to a third party, such as a DM. This can be used to distinguish a rightful owner from a pirate that may attempt to monetize a pirated dataset in a DM. If the DM, or the rightful owner detects such an event, the dataset can  be removed and the pirate be banned. This would mimic what web-sites like YouTube do to protect copyrighted content. Detecting the presence of pirated copies can be achieved using content similarity~\cite{ibosiola2018movie}, locality sensitive hashing~\cite{8003392,localhashing20}, and even hashing similarity~\cite{chang2019fbhash} that go beyond the scope of watermarking.

In addition to proving ownership, our watermarking technique can also reveal who may have leaked (copied/pirated) a dataset in the first place. A dataset seller or a DM may create a different watermark for every buyer and in addition to encoding it into the data, store also a description of it in some immutable index (e.g., a blockchain). Then, if an unauthorized copy of the dataset is found at a latter point, the culprit can be identified by looking up its watermark against this index.  

\noindent \textbf{Our major contributions} are as follows:  
\noindent $\bullet$ Our first contribution is the idea of using the appearance frequency of tokens to encode invisible watermarks upon datasets traded in DMs. We establish a family of such watermarks using frequency pairs and modulo arithmetic and prove that creating an optimal \textit{FreqyWM} reduces to solving a Maximum Weighted Matching (MWM) problem \cite{osiakwan1990maximum,galil1986maxwm} combined with a polynomial special version of the 0/1 Knapsack problem \cite{cormen2009introduction} involving items of equal value but different weights.

\noindent $\bullet$ We extend frequency-based watermarking to make it resilient against a series of attacks. In particular, we protect our technique against a \emph{Guess Attack} attempting to identify our watermarked pairs and secrets to impersonate the rightful owner. We make such an attack computationally hard by introducing a high-entropy secret while generating the watermark. We also protect against a \emph{Re-watermarking Attack} mounted by having a pirate inject its own watermark upon an already watermarked dataset, and then present the former as a false proof of ownership. We thwart such an attack by describing a simple protocol capable of ordering chronologically multiple watermarks that may be carried by different versions of the same dataset. We protect against a \emph{Destroy Attack} attempting to destroy our watermark by changing the frequency of different tokens in the dataset. By relaxing our modulo arithmetic rule used during the verification of a particular watermark pair, as well as the percentage of pairs to be detected before the entire watermark is verified (accepted), we oblige the attacker to effectively also destroy the actual data in the process of destroying the watermark. 
Finally, we show that our technique is robust to a \emph{Sampling Attack} in which the attacker attempts to pirate only a random sample of the watermarked data.

\noindent $\bullet$ Our final contribution is an extensive performance evaluation study aiming to explain the impact of the main parameters of \textit{FreqyWM} on major performance metrics under different attack scenarios using synthetic and real world datasets.  

\noindent The main \textbf{findings} of our evaluation are as follows: 

\noindent $\bullet$ We show that as long as there exists sufficient variation in the frequencies of different tokens, \textit{FreqyWM} can encode robust watermarks with minimal distortion on the initial data. 
Our technique does not apply to uniform token appearance frequencies, because in this case there does not exist sufficient gap between different frequencies for encoding a watermark. 

\noindent $\bullet$ Regarding the false positive probability, i.e., ``detecting'' a watermark on a dataset that does not carry it, our analytical bounds (in the form of closed form expressions) show that it quickly goes to zero as we increase the number of pairs. 

\noindent $\bullet$ We demonstrate that a Guess Attack has negligible probability of success, thereby making it impossible for almost all practical cases. On the up side, the rightful owner or any party, that is given the watermarking secret for verifying the watermark, can do that very fast in linear time complexity. %\item

\noindent $\bullet$ Regarding Sampling Attacks, we show that with the exception of very small samples, our detection algorithm is capable of detecting our watermark. Achieving this requires using the relaxed detection algorithm that trades robustness to attacks with false positives. For example, on a sample of $20\%$ and with thresholds that impose tiny false positive probability, the detection probability exceeds 90\%.%  \item 

\noindent $\bullet$ In terms of Destroy Attacks, we show that a watermark that imposes (costs) a tiny $0.0002\%$ distortion on the original data, remains detectable even under attacks that add random noise that imposes a $90\%$ modification. 

\noindent $\bullet$ Compared to existing solutions from the literature~\cite{shehab2007watermarking,dbwmli22} that are applicable only to numerical data and preserve only the mean value of the watermarked data, \textit{FreqyWM} allows a data owner to control the exact amount of distortion introduced by the watermark in terms of cosine or other similarity metrics which, under ~\cite{shehab2007watermarking,dbwmli22} may become unbounded. For example, a \textit{FreqyWM} watermark that imposes only $0.0002\%$ distortion in terms of cosine similarity, is stronger than watermarks from \cite{shehab2007watermarking} and \cite{dbwmli22} that impose $46.72\%$  and $4\%$ distortion, respectively under the same metric.

\section{Related Work}
\label{relatedwork}

Database watermarking is the closest type of watermarking to our work. There are of course other types of watermarking and fingerprinting  (when an owner generates a unique watermark for each intended party, e.g., buyers/data marketplaces), for example, for sequential  \cite{ayday2019robust} and genomic datasets \cite{ji2022robust}.
However, as they focus on specialized types of data, we do not go into more details about them.
Survey papers such as~\cite{kumar2020recent,kamran2018comprehensive,panah2016properties,RaniH22} compare database watermarking techniques in terms of verifiability, distortion, supported data types, and other aspects. Many of these solutions are applicable only to numerical data and thus cannot be applied to a range of commercial datasets, e.g., to web-browsing click-streams. 

The first known watermarking technique for relational data is a \textit{numerical database watermarking} approach \cite{agrawal2002watermarking}.
The watermark information is normally embedded in the Least Significant Bit (LSB) of features of relational databases to minimize distortion. 
Other numerical database watermarking solutions introduce distortion by considering the statistics of numeric values \cite{jony2021robust,shehab2007watermarking,farfoura2012blind,li2006publicly,hu2018new}. The proposed solutions in \cite{agrawal2002watermarking,shehab2007watermarking} focus on keeping the change at minimum (i.e., median and standard deviation). 
However, numerical database watermarking unfortunately cannot be applied to datasets composed of string and numerical values (e.g., CDRs, web-browsing history) that we handle in our work.

\textit{Distortion-free database watermarking} schemes have also been proposed \cite{el2010novel,darwishdistortion} that introduce fake tuples or columns in the original database. The fake tuples or columns are created based on a watermark secret by computing a secret function which makes watermarking visible and easy to remove.
\textit{However, an attacker can remove the watermark with minimum computational power}, making these approaches inapplicable to our case. 
\textit{Reversible watermarking} allows owners to reconstruct the original data used for watermarking on the top of detection
\cite{bigdb21,dbwmli22,zhang2008reversible,wang2022fbipt,gupta2009reversible,jawad2013genetic,imamoglu2017new,hu2018new,chang2021reversible}. 
They have similar properties as other relational watermarking techniques (e.g., private key based, robust, introducing distortion).

\textit{Categorical watermarking} \cite{categorywm21} is another watermarking approach that replaces tokens in a dataset with another token in the same category. However, this causes an undesired distortion and requires predefined categories (e.g., gender, clothing size) in the data. Consequently, its applicability on datasets consisting of different data types is limited.
\textit{Text watermarking} \cite{dbwmli22,textwm2021} is for text files where it changes a token (e.g., by replacing a word with another similar word) trying to preserve the meaning of a text \cite{dbwmli22} and/or the frequencies of the words \cite{textwm2021}. 
However, assume the dataset is a list of URLs visited by the owner, then this (insecure) change/replacement may invalidate a token (e.g., causing an invalid URL).

In the context of datasets in our use case, while prior works try to minimize the amount of distortion on median, average, or first moments of the distribution of a feature, the owner can limit the exact distortion between the original and the watermarked dataset as reflected by distance metrics that capture the shape of the entire distribution of a feature. Our results in Section \ref{comparison} have shown that the latter can deviate arbitrarily if an owner tries to control only the first most important moments.

\section{Frequency-based Watermarking}\label{freqy}
In this section we provide an overview  of \emph{FreqyWM} and the notations used throughout the paper in Table \ref{tab:notation}.
\begin{table}[htp]
	\centering
	 \resizebox{0.9\columnwidth}{!}{
		\begin{tabular}{|c|c|}
			\hline 
			$D_o$ &  The original data to watermark.\\
			\hline
			$D_w$ &  Watermarked (data) version of $D_o$.\\
			\hline
			$\uL_i$ &  $i$th \uP. \\
			\hline
			$f^o_i$ &  Frequency of $i$th \uP \ in $D_o$. \\
			\hline
			$f^w_i$ &  Frequency of $i$th \uP \ in $D_w$. \\
			\hline
			$R$& A high entropy secret. \\
			\hline 
			$L_{wm}$& A list of chosen \uP \ pairs for watermarking. \\
			\hline 
			$L_{sc}$& A list of secrets required for watermark detection. \\
			\hline 
			$L_{e}$& A list of eligible token pairs for watermarking. \\
			\hline
			$k$ & Threshold for detecting a watermark. \\
			\hline 
			$t$ & Threshold to accept a pair as watermarked. 
			\\ \hline
			$b$ & A budget threshold for distortion that watermarking can introduce. \\ \hline
	\end{tabular}}
    \caption{Notation.}
	\label{tab:notation}
\end{table}

\begin{figure*}[htp]
  \centering
\includegraphics[width=0.9\textwidth]{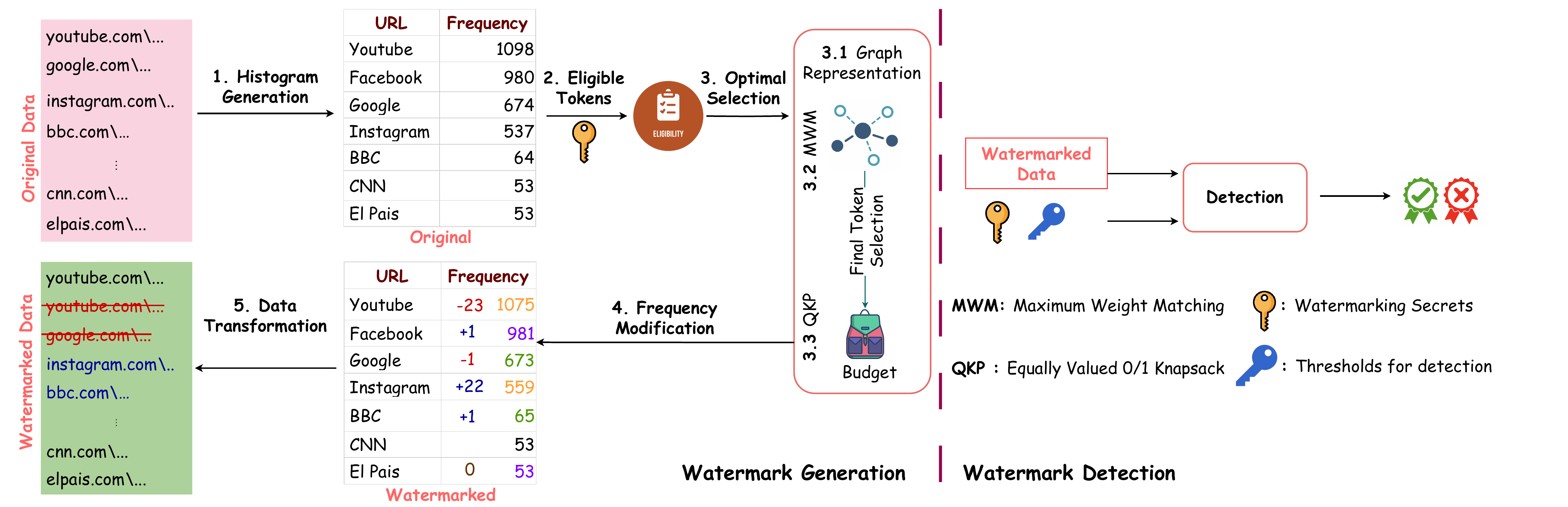}
 \caption{{\small \textit{FreqyWM}  illustrated based on a (Top Level Domain, TLD) URL dataset. URLs chosen as a pair for watermarking are represented with the same colored frequencies (e.g., \texttt{Youtube} and \texttt{Instagram}) while the ones not selected are colored black (e.g., \texttt{CNN}).
}}
 \label{fig:overview}
\end{figure*}

\textit{Running Example.} To provide the intuition behind our watermarking approach, assume a scenario where an owner holds a real click-stream dataset consisting of visited URLs (e.g., the dataset by \cite{iordanou2019beyond}).
Such datasets are desired by modern data analytic-based applications \cite{frequencyvldb21} where their frequency histograms (e.g., the number of clicks/visits, popularity of likes in social networks) are used as an essential source of information. 
For instance, assuming the appearance frequencies (histogram) visualized in Figure \ref{fig:overview} via a tabular form, the most frequent token is \texttt{youtube.com}, the second one is \texttt{facebook.com}, and so forth. 
After watermarking, it is important that the \textit{ranking} of the tokens based on the frequency shall not change while the frequency appearances can be modified. 
For instance \texttt{youtube.com} shall be the most frequent URL (token) visited in the watermarked dataset.
Another important distortion metric on the histogram is \textit{similarity}. 
It is important that owners shall have control over the change in similarity.
Since the similarity metric can be varied depending on the application that a dataset will be used, owners can assign a \textit{budget} to determine the minimum similarity desired on the frequency distribution after watermarking. 
Based on the above, we derive two natural constraints on the data utility to allow an owner to control distortion, without limiting watermarking to a specific data type: 

\noindent $\bullet$ \textit{Ranking Constraint:} Watermarking should preserve the ranking of token frequency distribution (histogram). Preservation of ranking does not of course imply that frequencies of individual tokens need to remain intact.

\noindent $\bullet$ \textit{Similarity Constraint: }The similarity between original and watermarked frequency distributions  (histograms)  should not be any less than $(100-b)\%$ where $b$ is a budget. Input $b$ is determined by the owner to keep distortion due to watermark generation within $b$. \footnote{Although in our experiments we use cosine similarity, any similarity metrics can be deployed without any loss of security and change in \textit{FreqyWM}.}  

To satisfy these constraints and overcome the shortcoming of existing watermarking techniques, we introduce a new \textit{private-key based} watermarking scheme, \textit{FreqyWM}, that is \textit{blind} (does not require the original data),  \textit{primary-key free} (\revised{does not need attributes that uniquely specify a tuple in a relation in a dataset}), \textit{robust}, and \textit{secure} against \textit{guess}, \textit{sampling}, \textit{destroy}, and \textit{false-claim} attacks with a high utility and a good trade-off between the complexity of the transformation and algorithmic efficiency of the solution.
\subsection{Overview of our Approach}
\textit{FreqyWM} consists of two main algorithms: the watermark generation algorithm, $WMGenerate$, and the watermark detection algorithm, $WMDetect$. $WMGenerate$ generates watermarked data based on a \emph{budget} $b$ capturing how much the watermarked data may differ from the original one, e.g., in terms of cosine (or other) similarity metrics of their corresponding token frequency distributions. By calling $WMGenerate$, the owner creates a watermarked version of their data consisting of \uP s such that ownership can be proved. $WMDetect$ detects if a suspected dataset holds the watermark of the owner using the owner secrets produced by $WMGenerate$ and two thresholds ($k$ and $t$). If $WMDetect$ outputs \textit{accept/verified}, this evidence would prove that the owner can claim ownership of the watermark and thus the data. By nature, $WMDetect$ can be computed as many times as desired in private while it can be computed \textit{only once} in public, because it would mean that the potential data owner shall reveal the secret leading to such watermark to the public (or whomever must verify it, e.g., a judging third party). As part of our future work, we are also looking at public verifiability without revealing the private key (Section~\ref{further}).

We describe the general idea behind \textit{FreqyWM}, illustrated in Figure \ref{fig:overview}. We use our running example. 
Of course, our technique is general and can be applied to any repeating token beyond just URLs, as we explain in Section~\ref{applicability}. \\
\textbf{Watermark Generation.} Assume that the data owner holding a list of URLs visited creates a dataset $D_o$ using the \textit{domain} of each URL in the list as a token 
and sets a budget $b$ for the similarity constraint. 
$WMGenerate$ has the following steps:

\noindent $\bullet$  \textit{Histogram Generation.}
Since \textit{FreqyWM} aims to preserve the appearance frequency of tokens, it first creates a histogram of the original dataset $D_o$ such that it sorts all unique tokens in descending order of their frequency (e.g., YouTube is the most visited, Facebook is the second, and so on).

\noindent $\bullet$  \textit{Generation of Eligible Tokens.}
 \textit{FreqyWM} cannot modify the frequencies randomly because of the ranking constraint. Therefore, it identifies a list $L_e$ of eligible pairs  of \uP s that are candidates to be watermarked using some secret $R$. 

\noindent $\bullet$  \textit{Optimal Selection}. 
 With the identification of eligible pairs, \textit{FreqyWM} ensures that the ranking is preserved after watermarking. 
However, the similarity constraint is yet to be satisfied. To keep the similarity at least at $(100-b)\%$, \textit{FreqyWM} selects pairs of \uP s from eligible pairs for watermarking, denoted by $L_{wm}$, based on the budget constraint $b$.
For this purpose, \textit{FreqyWM} benefits from solving two well-known problems: \textit{Maximum Weight Matching} (MWM) and \textit{Equally Valued 0/1 Knapsack problem} (QKP). 
To do so, eligible pairs are converted to a graph representation where vertices represent a token, and an edge represents a pair. 
\textit{FreqyWM} applies Maximum Weight Matching to the graph representation (discussed in detail later). 
By applying MWM, \textit{FreqyWM} selects the pairs from eligible pairs requiring the minimum change in total; however, it does not necessarily mean that the similarity between the original histogram and watermarked histogram will be at least $(100-b)\%$.
To choose another set of pairs satisfying the Ranking Constraint from the pairs derived after MWM, an \textit{Equally Valued 0/1 Knapsack problem} needs to be solved. 
The more the token pairs are selected to watermark, the more robust \textit{FreqyWM} is, since the number of tokens to attack (e.g., remove/identify) increases.
To fulfill the budget $b$, QKP selects a maximum number of pairs such that the similarity between the original frequency histogram and the watermarked one is \emph{at least} $(100-b)\%$. 

\noindent $\bullet$ \textit{Frequency Modification}. Until now, \textit{FreqyWM} determines the final pairs of tokens for watermarking but frequency appearances are yet to be modified to create the watermarked histogram. 
Therefore, \textit{FreqyWM} modifies the frequencies of the selected \uP s where the frequencies of a pair of \uP s would be equal to $0$ (as a watermark embedding rule) in some modulo that is calculated based on secrets and \uP s in the pair. 
To make it more comprehensible and show how the modifications occur, let us assume that the frequencies of a chosen pair, e.g., \texttt{youtube.com} and \texttt{instagram.com}, are $1098$ and $537$, respectively.  
Assume also that a modulo value, say $129$, is computed based on the secrets and the tokens (e.g., \texttt{youtube.com} and \texttt{instagram.com}). 
The difference between the two frequencies in modulo $129$ is $45$. 
To set the difference to $0$, we need to change the appearance frequencies for Youtube and Instagram in the dataset. 
$45$ is divided (by $2$) as $23$ (by ceiling) and $22$ (by flooring). 
The new frequencies of \texttt{youtube.com} and \texttt{instagram.com} need to become $1098-23=1075$ and $537+22=559$ such that $(1075-559) \mod 129 \equiv 0$. 
We can do that by removing $23$ instances of Youtube from the dataset, and adding $22$ more instances of Instagram. 
However, when the remainder (i.e., $(1098-537) \mod 21 \equiv 16$) is greater than half of the modulo, we add the modulo result calculated as $(\lceil(1098-537) \div 21 \rceil ) \times (1098-537)$ to the difference. 
This way, we never have to eliminate remainders that exceed half of the modulo. 
As it will be evident in the next section, this observation enables us to determine eligible tokens.\\%\item
\noindent $\bullet$ \textit{Data Transformation.} \textit{FreqyWM} adds/removes \uP s based on the frequencies and produces a watermarked dataset $D_w$. 
%\vspace{-3pt} 
\textbf{Watermark Detection.} An owner wishes to verify if a (watermarked) dataset $D'_w$ (a modified version of $D_w$) is watermarked by  using the secrets stored from the watermark generation. 
To determine the confidence level in the detection (e.g., the minimum number of detected watermarked tokens), the owner provides some threshold values  ($k$ and $t$).
With the watermarking secret and the thresholds, the detection returns \textit{accept/verified} or \textit{reject}.
\begin{freqywm}[label=WMInsert]{Watermark Generation}
\begin{small}
  \KwIn{$D_o,b$}
  \KwOut{$D_w, L_{sc}$} 
	$D^{hist}_o=\textbf{Preprocess}(D_o)$\;
	
	$R \gets \{0,1\}^\lambda $ , $ z \leftarrow Z^+$\; 
	
	\ForEach{$\{\uL_i,\uL_j\}_{i\ne j} \in D_o$}{$s_{{ij}}=H(\uL_i ||H(R ||\uL_j)) \mod z$\;}
	$L_e \leftarrow  \textbf{Eligible}(D^{hist}_o,\{s_{ij}\})$\; 
	
	$L_{wm} \leftarrow \textbf{OptMatch}(D^{hist}_o,L_e,\{s_{ij}\},b)$\; 
	
	\ForEach{$\{\uL_i,\uL_j\} \in L_{wm}$}{$D^{hist}_o.$\textbf{Update}($f^w_i,f^w_j,s_{ij}$)\; }%s.t. $(f^w_i -  f^w_j) \mod s_{ij} \equiv 0$}
	$D_w \leftarrow \textbf{Create}(D^{hist}_o, D_o)$\; 
	
	$L_{sc}=\{L_{wm},R,z\}$\; 
	
	\KwResult{$D_w,L_{sc}=\{L_{wm},R,z\}$}
\end{small}
\end{freqywm}
\subsection{Detailed Description of \textit{FreqyWM}}\label{frewywmdetail}
\subsubsection{Watermark Generation}
The data owner holds the original data $D_o$ and defines a budget $b$ that decides how much distortion a watermark can introduce. 
For comprehensibility, assume that $D_o$ is a single-dimensional dataset, e.g., a dataset with one attribute (see Section \ref{applicability} for how to apply \emph{FreqyWM} to multi-dimensional datasets).
$D_o$ consists of repeating values called \emph{tokens} that can be of \emph{any} data type, which enables \emph{FreqyWM} to be data-type agnostic.
The goal of watermarking is to generate the optimal watermark, i.e., \emph{with the largest number of watermarked pairs} within the given budget $b$.  
The generation algorithm (Algorithm \ref{WMInsert}) follows these steps:

\noindent \textbf{Histogram Generation}:  It pre-processes $D_o$ to  generate a histogram $D^{hist}_o=\textbf{Preprocess}(D_o)$.
$D^{hist}_o$ consists of a set of tokens, $\{ \uL_0,\dots, \uL_{|D^{hist}_o|}\}$ (e.g., $\uL_0=$\texttt{youtube.com}) where each $\uL_i$ has an (original) appearance frequency $f^o_i$ (e.g., there are $1098$ YouTube visits). 
The histogram $D^{hist}_o$ is sorted in a descending order of frequency. 
To keep the distortion introduced by the watermark at minimum (e.g., after watermarking, YouTube is still the most visited, followed by Facebook, although  their frequencies may have changed), we calculate two boundaries for each token $\uL_i$: an upper boundary $u_i$ and a lower boundary $l_i$. 
The boundaries allow us to determine how much change we can introduce to the token and whether a token pair is eligible as explained later. 
Naturally, for the \uP \ with the highest frequency in the histogram, it is $u_0=\infty$ because we can increase the frequency of $\uL_0$ as much as we want, while the lower boundary of the last token, $\uL_{|D^{hist}_o|-1}$, is set to its frequency as $l_{|D^{hist}_o|-1}=f_{|D^{hist}_o|-1}$ because we can remove at so many appearances. 
For the rest of the boundary calculations of each token $\uL_i$, $u_i$ is defined as the difference between $f^o_{(i-1)}$ and $f^o_i$, while $l_i$ is assigned as $f^o_i - f^o_{(i+1)}$. 
Note that once the boundaries are set, they remain same until frequency modification.\footnote{\revised{The frequencies of some tokens may have high importance. An owner can filter the dataset and exclude them from watermarking.}}

\noindent \textbf{Generation of Eligible Tokens}:  In cryptography,  $\lambda \in \mathbb{N}$ is a security parameter, i.e., a variable measuring the probability with which an adversary can break a cryptographic scheme \cite{KatzLindell2014}. In other words,  $\lambda$ provides a way of measuring how difficult it is for an adversary to break a cryptographic scheme. \textit{FreqyWM} requires randomization to be secure by ensuring that an attacker has only negligible advantage to recover the watermark and create collision for false claim (e.g., coming up with another watermarking secret which returns accept on data not watermarked by it). Thus, we choose a hash function to overcome the collision.
In detail, a hash function $H$ (chosen from a family of such functions) is a deterministic function from an arbitrary size input to a fixed size output, denoted $H:\{0,1\}^*\rightarrow \{0,1\}^\lambda$. The hash function \cite{KatzLindell2014} is \textit{collision resistant} if it is hard to find two different inputs $m_0 \neq m_1$ that hash to the same output $H(m_0) = H(m_1)$.

Based on the above, to determine token pairs for watermarking, \emph{FreqyWM} first generates a high entropy random number, i.e., secret, $R\leftarrow\{0,1\}^\lambda$ and an integer $z \in \mathbb{Z}^+$. 
Then, it uses $R$ and $z$ to compute $s_{ij}$ values for modulo operation as: $s_{ij}=H(\uL_i ||H(R ||\uL_j)) \mod z$, where $||$ denotes concatenation. 
A set $L_e$ of all eligible pairs is generated by an algorithm $Eligible$ based on given pairs $\{\uL_i,\uL_j\}$ and corresponding $s_{ij}$ values as $L_e \leftarrow  Eligible(D^{hist}_o,\{s_{ij}\})$ . 
A pair is accepted as eligible if it satisfies that the boundaries of each token in the pair are at least $ \lceil s_{ij}/2 \rceil$ where $s_{ij} \geq 2$. 
$s_{ij}$ cannot be $0$ or $1$ because of modulo operation since modulo $0$ is undefined and modulo $1$ is $0$. 
Note that the size of $L_e$ is bounded by $[0,\binom{|D^{hist}_o|}{2}]$ where $0$ means that there is \emph{no eligible pair} while $\binom{|D^{hist}_o|}{2}$ means that \emph{all the possible pairs} of tokens are eligible.
After the eligible pairs are constituted, the boundary check is not necessary anymore since whichever set of pairs (that does not have a common token among) is chosen, the ranking will be preserved.

\noindent  \textbf{Optimal Selection}:  The eligible pairs are defined by ensuring the ranking constraint. 
However, to determine which subset of eligible pairs shall be selected such that chosen optimal number of pairs of tokens, denoted by a set $L_{wm}$, respect the budget constraint, it runs optimal matching algorithm from the eligible pairs $L_e$ using the frequencies and $s_{ij}$ values as $L_{wm} \leftarrow OptMatch(D^{hist}_o,$ $L_e,\{s_{ij}\},b)$.
In Section \ref{heuristics}, we show that for our optimal selection solution, we acutely reduce our problem to \textit{Maximum Weight Matching (MWM)} and \textit{Equally Valued 0/1 Knapsack problem (QKP)} problems to solve. We also devise two heuristics: \textit{greedy} and \textit{random}.

\noindent \textbf{Frequency Modification}: Based on $L_{wm}$, the algorithm  creates new frequencies of \uP s chosen from the optimal matching algorithm $(f^w_i - f^w_j) \mod s_{ij} \equiv 0$. 
This, of course, changes the boundaries of tokens; however, we do not need to update the boundaries as they are not needed anymore.

\noindent \textbf{Data Transformation}:  It generates or removes \uP s based on new frequencies. Note that the position of where to add \uP s is important for security of \textit{FreqyWM} against guess attack. 
Therefore, new \uP s should be added in random positions (see Section \ref{applicability} for more discussion). As a final step, it returns the list of \uP s $D_w$ and stores $L_{wm}$, $z$ value, and the random value $R$ as a list $L_{sc}$. %\\

\subsubsection{Optimal and Heuristic Approaches } \label{heuristics}
Given that all watermarked pairs have equal value in terms of proving ownership of the data, an optimal watermark is just a watermark of maximum size in terms of watermarked pairs, within the defined constraints (\emph{similarity} and \emph{ranking}). 

\textit{Optimal Matching}. Let us now define our optimal watermarking.
Let $G=\{V,E\}$ be a connected undirected graph which is the representation of frequencies driven from eligible pairs $L_e$. 
$V=\{v_1,v_2,\dots,v_{|V|}\}$ where $v_i$ represents $\uL_i$ and $E=\{e_1,e_2,\dots,e_{|E|}\}$ where $e(v_i,v_j)$ is the edge between $v_i$ and $v_j$. 
The weight of an edge $e(v_i,v_j)$, $w(e_i)$, is equal to $T- ((f^o_i-f^o_j) \mod s_{ij})$ where $T$ is a big value (e.g., $T>C$ where $C$ is the highest difference between two frequencies in the eligible pairs). 
Then, our optimal watermarking problem reduces to finding the maximum number of edges (pairs) such that \emph{no edge} has a common vertex and $b$ is not exceeded. 

\begin{definition}[Optimal Watermarking]
Let $OptWM(G(V,E),b)$ be the optimal watermarking with a budget of $b$ among an eligible set of items $L_e$ represented as a connected undirected graph $G(V,E)$. 
The optimal watermarking produces the maximum number of edges (pairs) while not exceeding the budget $b$ defined below:

{\footnotesize
\begin{equation*}
  MAX \ |M^w| , \enspace M^w=\{e_1,...,e_{|M|}\} \enspace s.t. \enspace sim(D^{hist}_o,D^{hist}_w) \geq (100-b) %\sum_{e_i \in L_e}w(e_i)\times x_i \leq b \quad x_i \in \{0,1\}.
\end{equation*}}
\label{ref:wmopt}
where $M^w$ denotes the chosen pairs for watermarking.
\end{definition}

The solution of the pairing problem is reduced to two well-known problems with polynomial time solutions: \textit{Maximum Weight Matching} (MWM) and \textit{Equally Valued 0/1 Knapsack problem} (QKP) (which we have a special case where all values are equal). Note that while the general $0/1$ Knapsack problem is known to be NP-Hard \cite{cormen2009introduction}, this special equally valued $0/1$ Knapsack problem would have a polynomial time (greedy) solution. 
In particular, our optimal pairing problem is reduced and solved as follows: 

\noindent $\bullet$ Find the maximum weight matching $M={e_1,e_2,\dots,e_{|M|}}$ as $M=MWM(G(V,E))$. 
Notice that $M$ includes the edges such that the sum gives the maximum weight. 
It refers to minimum weight for us since the weights are defined as $T- (f_i-f_{j} \mod s_{ij} )$ which makes the highest frequency difference have the smallest weight and the smallest one have the highest weight. 
With this conversion, we identify the edges distorting the histogram minimally. \\
\noindent $\bullet$ After finding the edges via MWM, we have one more constraint which is the budget $b$. 
The matching algorithm has to return the maximum number of matchings for which the distortion (e.g., based on cosine similarity) does not exceed $b$ which can be solved via QKP where the value of each item is $1$, and the weight is recomputed as $T-w(e_i)$. 
Recomputation is necessary because for the QKP we want to add as many items as possible that will be bounded by $b$. 
Therefore, it finds the set of edges $L_{wm}$ in $M$ such that the selected edges do not exceed the budget $b$ by employing the QKP as $L_{wm}=QKP(M,b)$ where $L_{wm}={e_1,e_2,\dots,e_{|L_{wm}|}}$ and value of each $e_i$ is $1$ ($val(e_i)=1$).
Showing the optimality of the resulting watermark according to Definition \ref{ref:wmopt} is straightforward and can be proven via proof-by-contradiction. In a nutshell, if our solution is not optimal, it means that one of the solutions produced by $MWM$ and $QKP$ cannot be optimal. However, since $MWM$ and $QKP$ are both assumed to be optimal, this contradicts with our statement and thus our solution is optimal.

\textit{Heuristic Matching Algorithms.} We define two heuristic algorithms: 1) \emph{greedy}; and 2) \emph{random}. 
In the \emph{greedy} algorithm, all the eligible token pairs are sorted in an ascending order by their remainders as $rm_{ij} \equiv (f^o_i - f^o_j) \mod s_{ij}$. 
The algorithm starts selecting a pair respectively for watermarking where  $b$ would not be exceeded when it is chosen (i.e., comparing the similarities of original and watermark histograms). 
This continues until $b$ is exhausted or there is no more item to visit. 
The \emph{random} matching algorithm follows the same steps as the greedy algorithm except it does not sort the eligible pairs but rather selects a pair randomly from $L_e$.
\subsubsection{Watermark Detection}\label{wmdetect}
In detection, the data owner wishes to know if there is a watermark of its in a token dataset $D'_w$ to claim ownership. 
The owner holds its secret list $L_{sc}=\{L_{wm},R,z\}$ where $L_{wm}$ is the list of watermarked token pairs, $R$ is the high entropy value, and $z$ is the (modulo) integer, all generated by the watermark generation, along with two thresholds: (1) $t$, a \textit{threshold} to decide if a certain pair is watermarked; and (2)
$k$, the \textit{minimum number of watermarked pairs} required to conclude whether $D'_w$ is a watermarked dataset. 
How to set $t$ and $k$ depends on the robustness an owner wants (see Sections \ref{falsepositive} and \ref{limitoft}).  
If the owner wants to prove ownership to a third party, it has to reveal its secrets to that party. This causes to prove the ownership once in public (see Section \ref{falseclaim}). 
Our watermark detection algorithm (Algorithm \ref{WMDetect}) proceeds as follows:

\noindent \textit{(1)} It builds the histogram list $D^{hist}_w$ of the suspected dataset $D'_w$ as in the watermark generation algorithm. The algorithm does not calculate the boundaries, just the  \uP~frequencies.\\ 
\noindent \textit{(2)} For each \uP \ pair $\{\uL_i,\uL_j\}$ in $L_{wm}$, if the pair exists in $D^{hist}_w$, the algorithm  generates $s_{ij}$ values as $s_{ij}= H(\uL_i ||H(R ||\uL_j)) \mod z$.\\%\item 
\noindent \textit{(3)}  Then, it decides whether it will accept a given token pair $(tk_i,tk_j)$, as watermarked or not by checking  if the following statement holds: $(f_{i} - f_{j}) \mod s_{ij} \leq t$.\\
\noindent \textit{(4)} After finding which pairs are watermarked, it checks whether their number is over the \emph{minimum number of pairs}, $k$, needed to conclude that $D'_w$ is watermarked by the owner, and returns $accept$ (verified) or $reject$, accordingly.

\subsubsection{Probabilistic Analysis of False Positives} \label{falsepositive}
We develop a statistical bound in the form of the closed form expression derived from Markov’s inequality theorem, to demonstrate that the false positive probability (i.e., accepting a dataset as watermarked when it is not) goes to zero if we increase the minimum number of pairs $k$ that has to be accepted, or if we decrease the threshold $t$ to accept a pair as watermarked. 
Recall that the $m$-th token pair $\{\uL_i,\uL_j\} \in L_{wm}$ is accepted as watermarked, if  $(f_{i} - f_{j}) \mod s_{ij} \leq t$. 
We represent the probability that this ``watermarking statement'' holds as $P(X_m=1)=p_m, \text{ for } m=1,...,n $.  
$p_m$ depends on $t$.
The logic is that if $t$ is zero, the false negatives will increase. 
Let us assume that $p_m$'s follow a \texttt{Uniform[0,1]} distribution.
\footnote{They can follow different distributions, such as Uniform, Normal, Beta, as Markov’s bound is a general property and it can be applied any distribution.} 
This way the probability of success for accepting a pair as watermarked pair is $p_{m}=\{|t|/s_{ij}\}$. 
The probability of having at least $k$ successes in $n$ trials can be written as $P(S_n \geq k) = \sum_{i=k}^n P(S_n = m)$. 
Recall that $k$ represents the minimum number of pairs that has to be ``accepted'', to say that the dataset $D'_w$ is watermarked by the owner.
We now study the behavior of $P(S_n \geq k)$ depending on the behavior of $t$ and $k$ by using the \textit{Sandwich Rule} and Markov's upper bound obtained by its inequality theorem $P(S_n \geq k) \leq \frac{\mu}{k}$, where $\mu=\sum_{m=1}^n p_m$ is the mean of $S_n$.
Our analysis shows that if we decrease $t$, the probability of accepting a dataset as watermarked goes to zero and if we increase $k$, it will be hard to ``accept'' a dataset as watermarked.

\noindent \textbf{What happens in the limit of $t$?} $t$ defines the robustness we want to have. 
If $t$ tends to zero, $p_m$ will tend to zero for all $m=1,...,n$, and then, $\mu$ will also tend to zero. Therefore, we shall see what happens when $\mu$ goes to zero. It is easy to see from Markov's bound that $ \lim_{\mu \rightarrow 0 } P(S_n \geq k) = 0 $. 
This means that if we decrease $t$, the probability of accepting a dataset as watermarked will decrease to zero.

\noindent \textbf{What happens in the limit of $k$?} $k$ defines the minimum number of ``accepted'' pairs of \uP s to conclude that the dataset is watermarked. 
If we increase $k$, it will be hard to ``accept'' a dataset as watermarked. 
Thus the probability of ``acceptance'' should be reduced. 
If $k$ tends to infinity, $ \lim_{k \rightarrow \infty } P(S_n \geq k)  = 0 $.
Let us see what happens if $k$ tends to $0$. 
From the CDF of a random variable, and since $S_n$ is non-negative, $ \lim_{k \rightarrow 0 }  P(S_n \geq k)  = 1 $. 
$P(S_n \geq k)$ against the value of $k$ (where $p_m$s are \texttt{Uniform[0,1]}), computed using the Discrete Fourier Transform  of the characteristic function of the Poisson-Binomial distribution for $n=50$. 
The survival probability is shown to be $0$ when $k$ goes to $50$.

\begin{small}
\begin{freqywm}[label=WMDetect]{Watermark Detection}
	\KwIn{$D'_w,L_{sc}=\{L_{wm},R,z\},k,t$}
	\KwOut{$accept/reject$}
	$D^{hist}_w=\textbf{Preprocess}(D'_w)$\;
    $count=0$, $result=reject$\;\\
	\ForEach{$\{\uL_i,\uL_j\} \in L_{wm}$}{
 \If{\textbf{Found($\uL_i,\uL_j$, $D^{hist}_w$)}}{
    $s_{ij}= H(\uL_i ||H(R ||\uL_j)) \mod z$\;
	
		\If{$(f_{i} - f_{j}) \mod s_{ij} \leq t$ }{
		   $count++$\;
		}}
}
	\If{$count \geq k$ }{
		 $result=accept$\;
		}
	\KwResult{result}%$\textbf{return }result$\;
\end{freqywm}
\end{small}

\section{Experimental Evaluation}
All of our experimental results are produced on a standard laptop machine with dual-core Intel Core(TM) i$7-5600$U CPU $2.5$GHz, $16.00$ GB RAM, $64$-bit OS, and implemented in Python language. 
We deployed SHA$256$ as a hash function. 
\subsection{Synthetic Experiments}\label{parameffect}

\begin{footnotesize}
    \begin{figure*}[htp]
\centering

\mbox{
\subfloat[{\small Effect of different skewness parameters ($\alpha$) on chosen pairs by \textit{Optimal}, \textit{Greedy}, and \textit{Random}.} ]{\includegraphics[width=.24\textwidth]{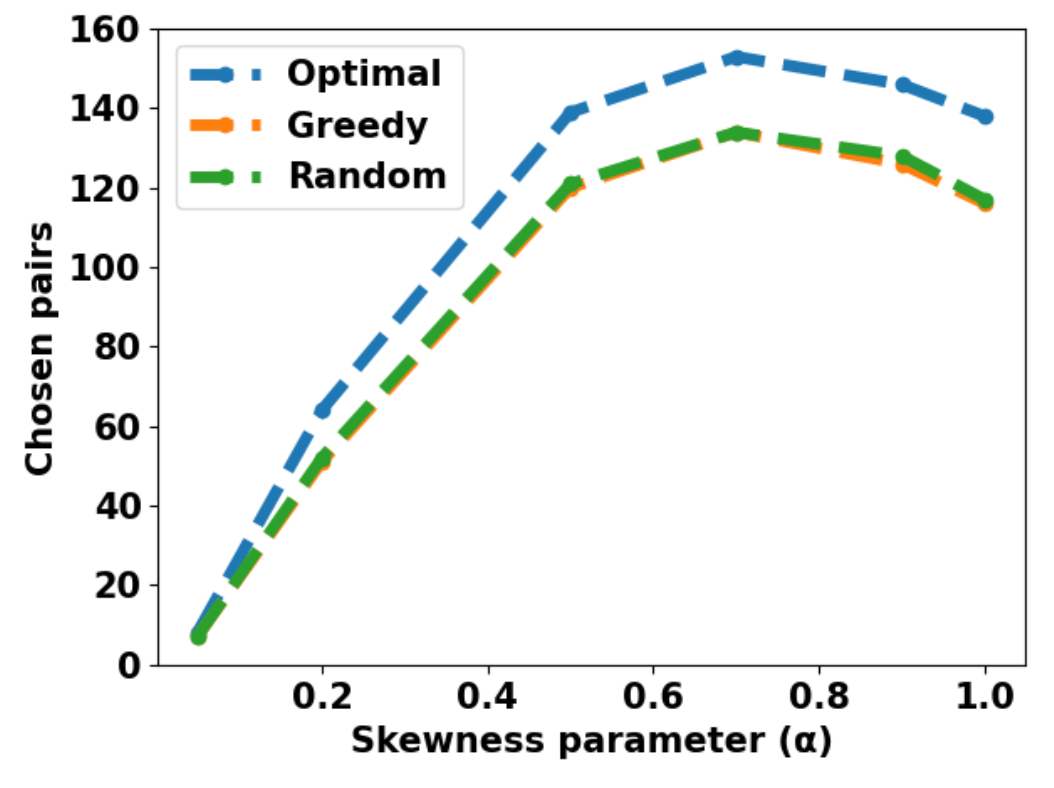}\label{fig:pairvsalpha} }\qquad

\subfloat[{\small Effect of different modulo values ($z$) on chosen pairs by \textit{Optimal}, \textit{Greedy}, and \textit{Random}.}]{\includegraphics[width=.24\textwidth]{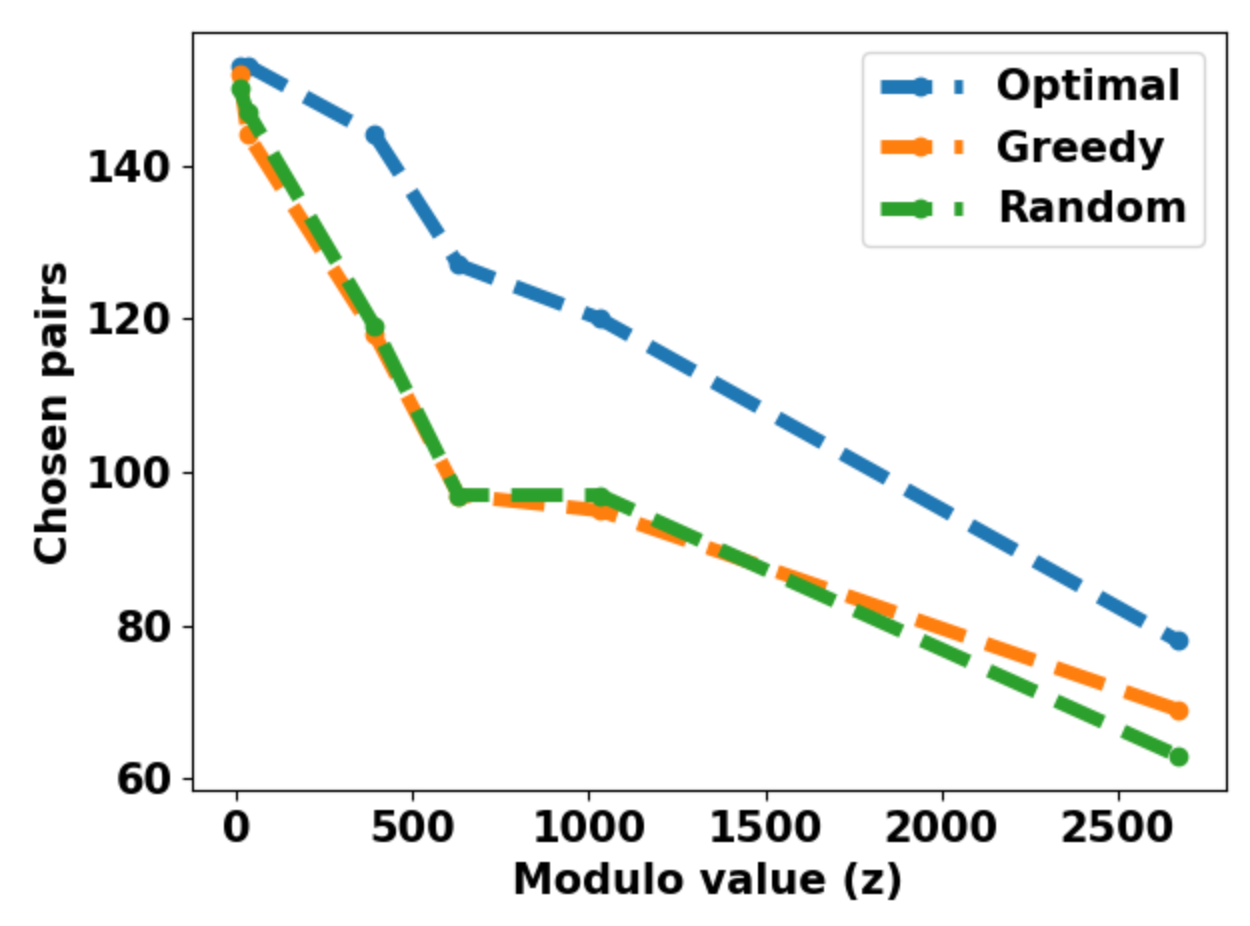}\label{fig:pairvsp}} \qquad

\subfloat[{\small Chosen pairs by \textit{Greedy} and \textit{Random} with respect to \textit{Optimal}.}]{\includegraphics[width=.24\textwidth]{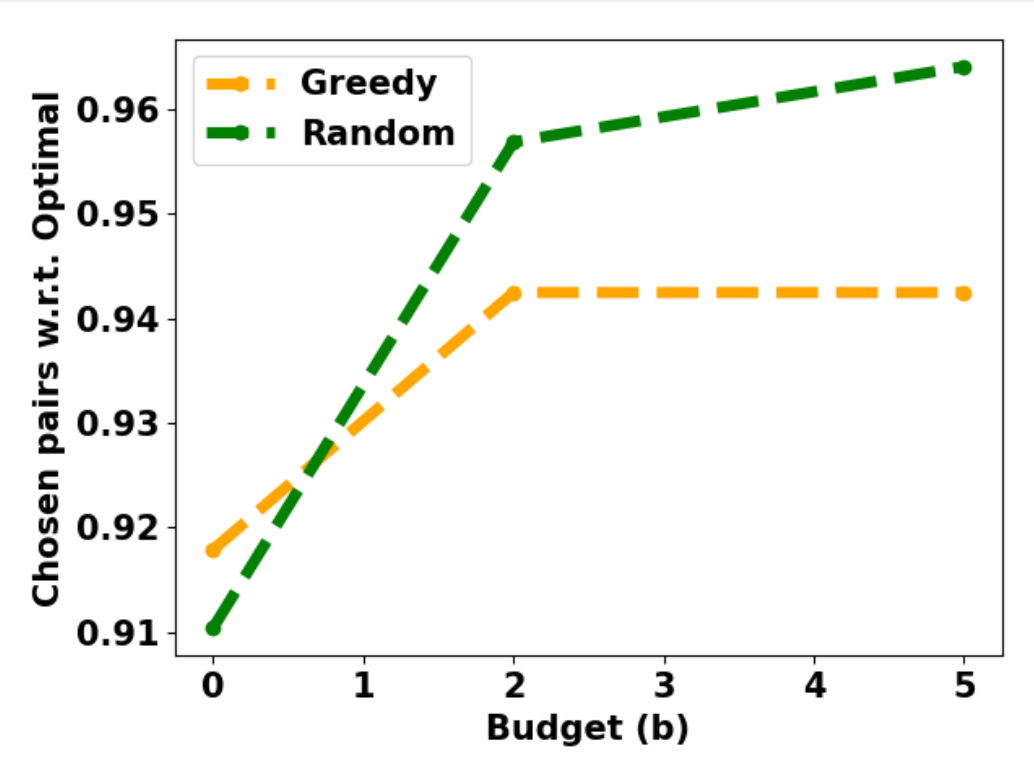}\label{fig:pairvsbudget}}
}
\caption{{\small Effects of parameters on the size of chosen pairs for watermarking.}} 
\label{fig:parameffect}
\end{figure*}
\end{footnotesize}

\begin{table*}[htp]
    \centering
 {\footnotesize
    \begin{tabular}{|c|c|c|c|c|c|c|c|c|c|}
    \hline
     \textbf{Dataset}  & \textbf{Size} &\textbf{Token} &\textbf{Distinct Tokens}& \textbf{$|L_e|$} & \textbf{Optimal} &\textbf{Greedy} & \textbf{Random}& \textbf{Gen}&\textbf{Detect}\\ 
      &  &&&  &  & & &(sec)&(sec)\\ \hline
     \textbf{Chicago Taxi} \cite{chicagotaxi} & $9.68$GB & \texttt{Taxi ID} &  $6573$ &  $33308$ & $805$ & $770$ & $773$& $182.51$& $ 0.609$\\ \hline
      \textbf{eyeWnder} \cite{iordanou2019beyond} & $247$MB & \texttt{URL} &  $11479$ &  $257$ &$38$ & $33$ & $31$  &$420.81$& $0.053$\\ \hline
       \textbf{Adult} \cite{adult} & $4$MB & \texttt{Age} &  $73$ &  $72$ &$21$ & $20$ & $17$  &$0.03$ & $0.001$\\ \hline
    \end{tabular}}
    \caption{{\small Validation results on real world datasets. \textbf{Dataset}: Dataset used  \textbf{Size}: The size of original dataset. \textbf{Token}: Definition of the token (e.g., the name(s) of the attributes). \textbf{Distinct Token}: The number of distinct tokens. \textbf{$|L_e|$}: The number of eligible pairs. \textbf{Optimal}: The number of chosen pairs by the optimal matching. \textbf{Greedy}: The number of chosen pairs by the greedy matching. \textbf{Random}: The number of chosen pairs by the random matching. \textbf{Gen}: Running time of watermark generation. \textbf{Detect}: Running time of watermark detection.}}
    \label{tab:realvalidation}
\end{table*}

For our synthetic experiments, we generated synthetic datasets using a $power-law$ distribution \cite{clauset2009power} with different skewness values $\alpha$ as $[0.05,0.2,0.5,0.7,0.9,1]$. The sample size is $1M$ and the number of \uP s is $1K$ for each different $\alpha$ value. When $\alpha$ is $0$, it is a uniform distribution in which there are no eligible \uP s to watermark. When $\alpha$ is $1$, the original dataset $D_o$ is skewed with a very long tail with almost equal values.  
In this setting, we evaluate how the parameters (a modulo value $z$, a budget $b$, and skewness parameter $\alpha$) are affecting the number of chosen pairs for watermarking and the performance of \textit{optimal}, \textit{greedy}, and \textit{random} approached in terms of number of chosen pairs. 

Figure \ref{fig:pairvsalpha} shows the correlation between skewness of a dataset $\alpha$ and the size of chosen pairs when budget $b=2$ and modulo value $z=1031$. When a dataset is almost uniform (i.e., $\alpha=0.05$), the solutions can select very few pairs since there are not many eligible items (i.e., the upper and lower boundaries are not enough, in fact many of them are $0$). When $\alpha$ starts increasing, the differences between the frequencies of tokens increase. Thus, the number of eligible items increases which also affects the number of chosen pairs under a given budget. However, at some point (i.e., $\alpha=0.7$), the number of chosen pairs decreases due to the tail of (histogram) frequencies becoming uniform. The same figure shows the superior performance of the optimal solution. The gap between optimal and the heuristics is around $20\%$ for most $\alpha$ values while the two heuristics perform similar the one with the other ($0.02\%$ in average). 

Figure \ref{fig:pairvsp} illustrates how the modulo value $z$ affects the size of chosen pairs. When we pick smaller modulo value $z$, we would have a higher number of chosen pairs. The reason is that a smaller $z$ leads to smaller remainders $s_{ij}$ that need to be eliminated, thereby yielding more selected pairs within a given budget $b$. When $z$ is very small (i.e., $10$), the three approaches are very close (also see Section \ref{guess} for the effect of $z$ in terms of security). However, when $z$ increases, our optimal approach selects many more pairs than the heuristics. 

Figure~\ref{fig:pairvsbudget} shows how the budget selection affects the performance comparison between the heuristics and the optimal. We set modulo value $z=1031$ and use the dataset with the skewness $\alpha=0.7$. When we increase the budget $b$, the heuristics get closer to the optimal performance. This is expected since even the optimal algorithm cannot select more than all the eligible pairs and with a large budget even the heuristics can approach that. 
\subsubsection{Limit of $z$}\label{limitofz}
We stated that $z$ is selected from $Z^+$; however, by analyzing the frequency histogram we can derive the upper and lower boundaries.
Since the minimum value (lower bound) $z$ can take is $2$, we delve into investigation of the upper bound of $z$.
Note that since the token with the highest frequency has the upper bound of infinity, there will be \textit{at least} one pair that could be used for watermarking. 
Assume a watermarking pair candidate $(\uL_i,\uL_j)$.
Their frequencies, $f_i$ and $f_j$, are changed such that $(f^w_i - f^w_j) \mod s_{ij} \equiv 0$. 
To have an upper bound for $z$, let us investigate which pair of tokens results in the highest difference. 
If we can determine the highest difference, say $r_{max}$, then $r_{max}$ can be assumed as the upper bound for $z$ since it is the highest remainder.
Now, considering $D^{hist}_o$, the highest difference is between $\uL_0$ (the token having the highest frequency) and $\uL_{|D^{hist}_o|-1}$ (the token having the lowest frequency).
That means that the largest remainder for any pair is $r_{max}=(f^0_0 - f^0_{|D^{hist}_o|-1})$.
Thus it is natural to accept that the upper bound of $z$ is $r_{max}$. 
To conclude, $z$ can be chosen from $(2,r_{max})$.
Overall, $r_{max}$ can be calculated as  $\forall f_i,f_j \in D^{hist}_o$ s.t. $f_i \geq f_j; \enspace r_{max}=\max(\{f_i-f_j\})$.
Hence, the upper bound for $z$ is calculated.
However, note that this value can be small and can be an advantage to an attacker.
As discussed in Section \ref{parameffect}, $z$ affects the number of chosen pairs; thus, it correlates with the mix of possible attacks and is use-case scenario dependent.
We plan to investigate this observation theoretically and experimentally in terms of security, robustness, and utility in the future.
\subsubsection{Limit of $t$}\label{limitoft}
Another critical parameter is $t \in [0,s_{ij}-1]$. Note that since $s_{ij}$ has an upper bound as $z-1$, the highest value assigned to $t$ is $z-1$.
While in our experimental study we chose $t$ as a constant value, $t$ could be also a percentage. 
Assume that an owner wishes to state that it wants $50\%$ of error tolerance. 
Now, setting $t=s_{ij}\times 0.5$ states that a pair, say $\uL_i$ and $ \uL_j$, will be accepted as a watermarked if $(f_{i} - f_{j}) \mod s_{ij} \leq s_{ij}/2$.
Thus, $t$ represents the robustness level an owner desires.
For instance, if $t=0$ then the watermark becomes fragile since it cannot tolerate any changes in $D_w$, thus missing watermarked pairs (i.e., high false negatives).
On the other hand, when $t=100$, it is more robust and can tolerate modifications in $D_w$; however, it also means that it accepts more false positives (see also Section \ref{falsepositive}). 

\subsection{Validation Using Real World Datasets} \label{datavalidation}
Next we apply \textit{FreqyWM} to three real world datasets from different domains: (1) \texttt{Chicago Taxi} dataset~\cite{chicagotaxi}; (2) A real click-stream dataset logging the URLs visited by a group of users of the \texttt{eyeWnder} advertisement detection add-on \cite{iordanou2019beyond}; (3) \texttt{Adult} dataset \cite{adult}. Our intention is to validate our main conclusion using real data from the previous evaluation with synthetic data, i.e., that the heuristic approaches perform well enough compared to the optimal. A second evaluation objective is to report the real processing time on an ordinary machine for generating and detecting the watermark using these real datasets.

For our watermark generation, we set the modulo value $z=131$ and the budget $b=2$. 
We run our algorithm $30$ times and take the mean of total computations. 
Table \ref{tab:realvalidation} presents our validation results. \texttt{Taxi ID}, \texttt{URL}, and \texttt{Age} were chosen as tokens for \texttt{Chicago Taxi}, \texttt{eyeWnder}, and \texttt{Adult}, respectively. After generation, for \texttt{Chicago Taxi}, \texttt{eyeWnder}, and \texttt{Adult} datasets, our optimal solution chose $805$, $38$, and $21$ pairs, respectively. Considering the heuristic approaches, greedy chose $770$ pairs for \texttt{Chicago Taxi}, $33$ pairs for \texttt{eyeWnder}, and $20$ pairs \texttt{Adult} while random chose $773$, $31$, and $17$ pairs, for \texttt{Chicago Taxi}, \texttt{eyeWnder}, and \texttt{Adult}, respectively. Running times of computations for watermark generation on the \texttt{Chicago Taxi}, \texttt{eyeWnder}, and \texttt{Adult} datasets were $182.51$ secs, $420.81$ secs, and $0.03$ secs, respectively (where we exclude histogram and watermarked data generations). For watermark detection, the total detection time for each watermarked datasets was less than $1$ sec. As it can be interpreted from Table \ref{tab:realvalidation}, the number of chosen pairs increases when the number of eligible pairs increases. For instance, while \texttt{eyeWnder} has more distinct tokens ($11479$) than \texttt{Chicago Taxi} has ($6573$), \texttt{eyeWnder} has fewer eligible pairs ($257$) than \texttt{Chicago Taxi} has ($33308$). Thus, \textit{FreqyWM} has selected more pairs for \texttt{Chicago Taxi} ($805$) than it selected for \texttt{eyeWnder} ($38$). 

\subsection{Watermarking Multi-Dimensional Data}\label{applicability}
During our discussion so far, we set the token as a single attribute.  
However, as we previously stated, a token does not necessarily need to be restricted to a single attribute of a multi-dimensional dataset. 
Therefore, a token can be also defined as combination of more than one attributes (e.g., \texttt{[Age, WorkClass]}) in the \texttt{Adult} dataset. 
We ran \textit{FreqyWM} on such token represented as \texttt{[Age, WorkClass]} with the same parameter setting in Section \ref{datavalidation} and the number of tokens (i.e., distinct \texttt{[Age, WorkClass]} attributes in the real dataset) were $481$. 
The size of pairs chosen for watermarking was $20$. With multi-dimensional data removing a token appearance is as simple as with single-dimensional data. 
Increasing, however, a token's frequency is more involved. 
The reason is that just repeating the value of the token would leave other fields not being part of the token with a value to be set, e.g., all the other fields beyond Age and WorkClass in the \texttt{Adult} dataset. 
A naive solution would be select a random appearance of the token and copy its other fields every time that an additional instance of the token needs to be added to the watermarked dataset. 
This, however, could create semantic inconsistencies if there are constraints to be respected for individual attributes or combinations of them. 
Making sure that added appearances of a token do not lead to semantic inconsistencies that, in addition to degrading the quality of the data, could also give away the existence of a watermarked pair to an attacker.
This analysis requires domain knowledge about what each dataset represents. 
Such knowledge, however, is orthogonal to all previous steps of our algorithms and, thus, can be appended as a last step based on one's domain knowledge of the data whenever a token's frequency needs to be increased. 
\revised{We are currently investigating them and the effect of \emph{FreqyWM} on data utility of such dataset with unique attributes as it is difficult to determine as addressed by~\cite{RaniH22}}.

\subsection{Comparison to Related Works}\label{comparison} 
As stated previously, we cannot directly apply (numerical) database watermarking to datasets similar to the ones we used for validation. However, one naive approach would be to convert a dataset to a numerical representation (e.g., a histogram) and watermark this numerical representation. In a nutshell, the histogram of a given dataset based on a predefined token is generated and then, the histogram is treated as a two dimensional database consisting of primary keys which are the tokens and an attribute consisting of integer values which are the frequencies. Later, a database watermarking is employed on this histogram. Then as in \textit{FreqyWM} data transformation (e.g., removing/adding tokens) occurs according to the (new) watermarked histogram computed by the database watermarking. However, applying a numerical database watermarking is not really straightforward since it will distort the underlying dataset unexpectedly as a result of change in histogram data (e.g., cosine similarity) and would require modification in their watermarking techniques (e.g., how to create a watermarked dataset from the watermarked numerical representation). However, since this is the closest and simplest approach, we compare against it. 

To actualize the above approach, we considered two numerical database watermarkings: 1) Shehab et al. \cite{shehab2007watermarking} (referred as WM-OBT) due to partitioning approach (i.e., grouping tokens before watermarking) similar to \emph{FreqyWM}; 2) Li et al. \cite{dbwmli22} (referred as WM-RVS) due to being one of the most recent reversible watermarking schemes introducing very small distortion compared to other same family of watermarkings.

More specifically, WM-OBT follows a data partition approach in which a watermark, defined as a bit sequence, is inserted on a group of partitions.
Each data partition is filled by tokens and the frequencies of the tokens in each partition are modified/distorted by solving a minimization (if a watermark bit is $0$) or maximization (if a watermark bit is $1$) problem via a genetic algorithm \cite{goldberg2007genetic}, in which the objective function is in the form of a sum of sigmoid functions. 
WM-RVS treats each numeric value individually and changes its decimal part by selecting the random least significant position based on the watermarking key/bit and attributes. 
Furthermore, to apply WM-OBT and WM-RVS on a histogram generated from a dataset, we had to adjust them such that their solutions produced are integers since a frequency count cannot be a decimal value.

\begin{figure}[htp]
    \centering
    \includegraphics[width=0.35\textwidth]{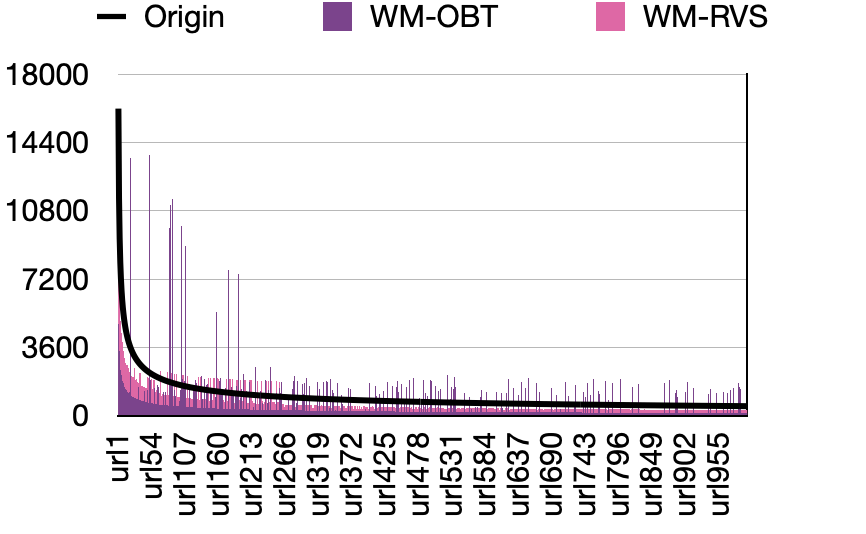}
    \caption{{\small Comparisons of the watermarked histograms generated from WM-OBT (purple color) and WM-RVS (fuchsia color) w.r.t. the original data histogram (black color) for the synthetic dataset with dummy token names.}}
    \label{fig:comparehist}
\end{figure}
For comparison, we investigate them based on two constraints: 1) change in the original histogram after watermarking (i.e., cosine similarity with watermarked histogram), and 2) the ranking of the tokens after watermarking. 

We ran \textit{FreqyWM}, WM-OBT, and WM-RVS on our synthetic data with skewness parameter $0.5$ (with $1K$ distinct tokens and $1M$ sample size) where we set $b=2$, and $z=131$ for \textit{FreqyWM}.
We set parameters for WM-OBT and WM-RVS such that the parameters are proportional to the experimental settings of Shehab et al.~\cite{shehab2007watermarking} and Li et al.~\cite{dbwmli22}. 
For WM-OBT, we use genetic algorithm (GA) technique for optimization ~\cite{goldberg2007genetic} where we fix the number of partitions as $20$ (where each partition has around $50$ tokens), watermark bit sequence as $[1,1,0,1,0]$, condition as $0.75$, and we allow the change (constraint) between $[-0.5,10]$.
The decoding threshold minimizing the probability of decoding error is calculated as $0.0966$. 
For WM-RVS, we use the same bit sequence as in WM-OBT without creating it from the chaotic encryption. 
Also, let us note that WM-OBT took more than $30$ minutes to run for such a small size dataset due to its optimization while WM-RVS was in the order of seconds.
Figure \ref{fig:comparehist} visualizes how the watermarked data histograms look like with respect to the original data histogram after applying WM-OBT and WM-RVS based on the experiments.

\noindent \textbf{Similarity.} In \textit{FreqyWM}, even with $2\%$ budget, the similarity between the original histogram and the watermarked histogram is $99.9998\%$, indicating that not all the budget was exhausted.
On the other hand, for WM-OBT and WM-RVS, the similarities are $54.28\%$ and $96\%$, respectively. 
The mean and standard deviation of the changes introduced to the histogram by WM-OBT are $444$ and $855.91$, respectively while they are $-69.43$ and $414.10$ for WM-RVS, respectively. 

\noindent \textbf{Ranking.} Another important evaluation is to compare the ranking of the tokens in the histograms under WM-OBT and \textit{FreqyWM}. 
\textit{FreqyWM} by definition maintains the ranking of tokens. 
Preserving the ranking allows us not to sacrifice the utility of a dataset, e.g., preserving the popularity of URLs. 
However, after our analysis, we observed that WM-OBT and WM-RVS changed the ranking of $998$ and $987$ out of the total 1000 tokens, respectively! 

The results on similarity and ranking support our claim that applying a numerical database technique on histogram data would result in unexpected and uncontrolled distortion that seriously undermines the utility of the original data. 

\section{Security and Robustness Analysis}\label{security}
This section discusses the security and robustness of our \textit{FreqyWM} method against four attacks: \textbf{ guess},\textbf{ sampling},\textbf{ destroy}, and \textbf{re-watermarking (false-claim)} attacks which are well-known attacks in watermarking as studied by ~\cite{RaniH22}.
In order to measure the robustness against sampling and destroy attacks, we run our optimal solution on a dataset where the skewness parameter $\alpha=0.5$ (with $1K$ distinct tokens and $1M$ sample size), unless stated otherwise, the modulo value $z=131$, and the budget $b=2$ and it selected $139$ pairs for watermark. 
We run the experiments for $100$ times and compute the average accepted pairs over all repetitions. 

\subsection{Guess (Brute-Force) Attack}\label{guess}
In the guess attack, the probabilistic polynomial time adversary tries to guess the watermark, i.e., the secret embedded in the data. This is possible only if it can figure out a subset of \uP \ pairs $\{\uL_i,\uL_j\}_l$ (where $\binom{|D^{hist}_w|}{2} \geq l \geq k $) based on the watermarked data $D_w$, the random value $R$, and the modulo value $z$ where the watermark detection algorithm based on these inputs (for some fixed $k$ and $t$) returns \texttt{accept}. Assuming that the hash function is collision resistant, $R$ is random, and $z$ is an integer, the probability of the attacker being successful can be formally defined as:  

\vspace{-6pt}
{\footnotesize
 \begin{align*}
    Pr[R\leftarrow \{0,1\}^\lambda; (D_w,L_{sc}= \{\{\uL_i,\uL_j\}_{|L_{wm}|},R,z\}) \\ \leftarrow  WmGenerate(D_o,b):\mathcal{A}(D_w)\rightarrow L'_{sc}=\{\{\uL_i,\uL_j\}_{l},R^*,z^*\} |\\ WmDetect(D_w,L'_{sc},k,t)=1] \leq \negl
\end{align*}
}
Considering the typical parameter values, the probability of success becomes negligible. 

\subsection{Sampling Attack}\label{sampling}
In this attack, $\mathcal{A}$ copies a random subsample from the watermarked dataset $D_{w}$ in an attempt to exploit (pirate/steal) it while hoping that the watermark won't be detectable within the extracted sample. 
The attack is run for different sample sizes from $1\%$ to $90\%$, extracted from the original watermarked dataset $D_{w}$.  
For each percentage and subsample we apply the detection algorithm and compute the percentage of accepted pairs.  
Also, for each subsample detection experiment, we deploy different values of the threshold $t$ for accepting a pair as watermarked as $t=\{0, 1, 2, 4, 10\}$.

The attack scenario is as follows: $\mathcal{A}$ randomly selects $x\%$ of $D_w$ where $x$ defines the percentage for the sampling attack (e.g., $1$) as a subsample size of $1M\times \frac{x}{100}$. 
When the owner suspects the dataset (possible subsampled), it scales it up to the size of $D_w$ by multiplying the frequency counts by $\frac{100}{x}$ by using its info from the (original) watermarked dataset (e.g., via info added to its metadata).  
For instance, for $1\%$ sampling attack, a subsample would have total of $1M\times0.01=10K$ where each $f_i$ is multiplied by approximately $0.01$. 
Note that if the sample size is greater than the number of distinct tokens, which is the number of items in $D^{hist}_w$, the sample will have all the distinct tokens with a high chance. 
This also means that all the chosen watermarked pairs are in the subsample. 
Our results show that the size of the extracted subsample does not greatly affect the number of accepted pairs if it is greater than the number of unique tokens ($1K$). 
Since the frequencies of the tokens vary, the value of $t$ does affect the result of the detection. 
For example, with $t=0$ the detection algorithm can detect around $36\%$ (in average) of the watermarked pairs. 
When $t$ increases from $1$ to $10$, the performance of the detection increases (in average) from $72\%$ to $99.5\%$.

\begin{figure}[htp]
\centering
\includegraphics[width=0.24\textwidth]{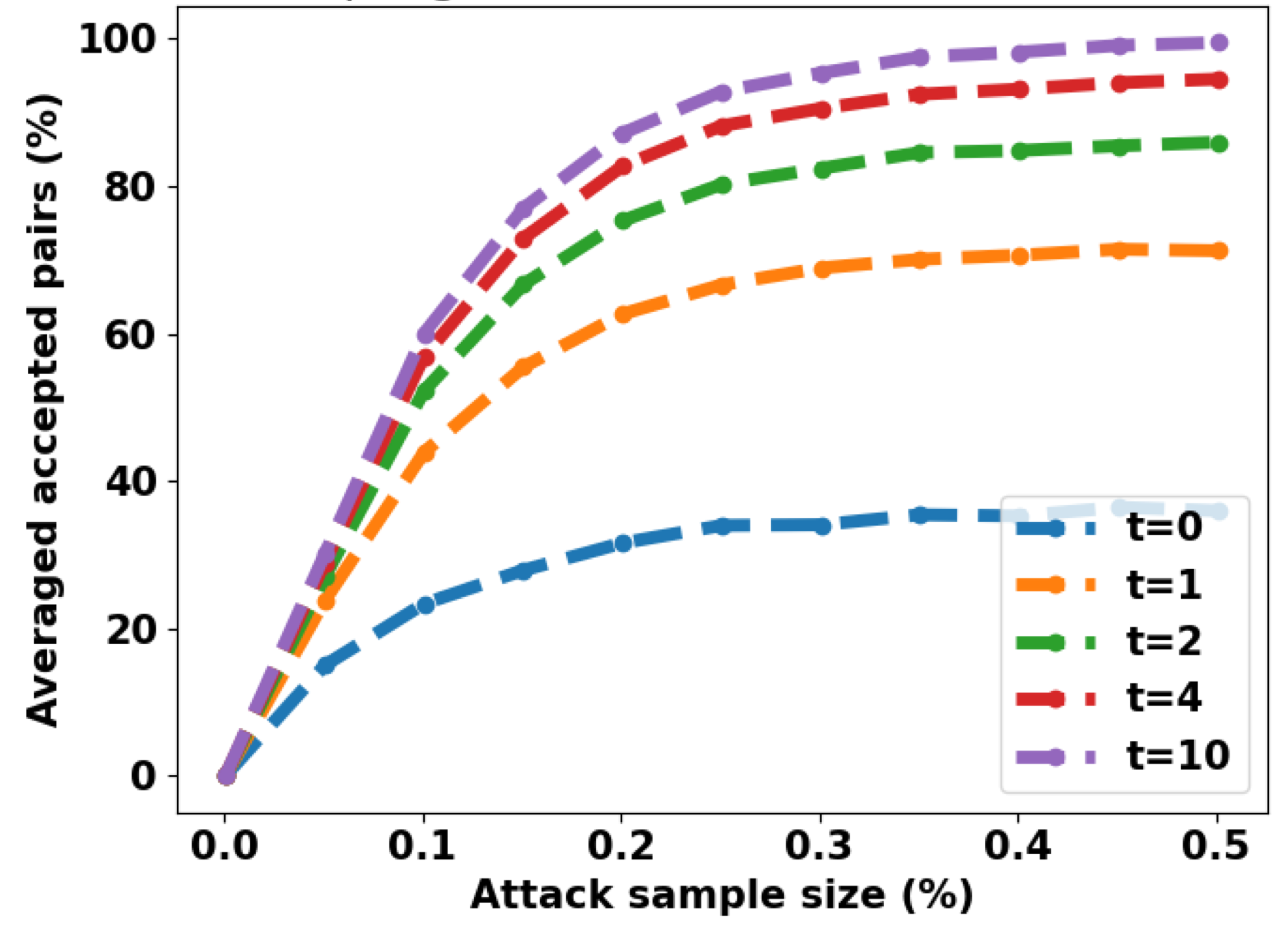}
\caption{{\small Sampling attack results with very low sample size and $\alpha=0.5$.}}
\label{fig:sampling05_verylow}
\end{figure}

Let us now see the results when the size of the extracted subsample is very low, so that it might not contain at least $1$ token from the total $1K$ of unique tokens that the original watermarked dataset has. 
Figure \ref{fig:sampling05_verylow} shows the results for sample size proportions between $0.0007\%$ and $0.5\%$. 
Observe that if the sample size is greater than $5\times$ the number of unique tokens ($1K$), the detection algorithm stabilizes its performance for detecting the watermark. 
Below $2\times$ ($2K$), the performance starts to decrease with higher velocity. 
In these extreme cases, the detection algorithm will have more difficulties to detect the data as watermarked. 
However, the utility of the data is highly degraded since the subsample sizes are very small compared to the original size of $1M$ tokens. 
This causes a small number of distinct tokens to be found in the subsample.

\noindent \textbf{Effect of modulo bases.} As seen previously, $t$ is crucial for detecting whether a pair is watermarked. 
For small values of $t$ to be sufficient to fend off sampling attacks, the remainders need to be small numbers that are covered by $t$. %the chosen
One way to achieve this is by ensuring that the modulo bases used (i.e., the $s_{ij}$'s) are relatively small numbers when compared to the actual appearance frequencies of watermarked pairs. 
When this does not apply, the method will of course fail. 
For instance, assume a watermarked pair involving frequencies $f_i=540, f_j=440$ which under base $s_{ij}=100$ leave a remainder of $0$. 
W.l.o.g, lets assume that a $50\%$ frequency attack leads to a dataset with $f_i=270, f_j=220$ which leads to a remainder of $(270-220) \mod 100 \equiv 50$. 
Now if $t$ is chosen smaller than $50$ then the watermarked pair will not be detected. 
The reason is that mod $100$ leaves large remainders when applied to $f_i$ and $f_j$ that are in the same order with $100$.
In our experimental results $f_i$'s were always much larger numbers than the employed $s_{ij}$, thereby, even small $t$'s would detect a pair under a sampling attack.
To determine the optimal $t$ and how robust it is against attacks, a further investigation is needed as it depends on various parameters such as $z, s_{ij}$ as well on the mix of expected attacks as discussed later. 
Note that our experimental results show that $s_{ij}$ values are $\sim2$ order of magnitude smaller than $z$. 
Furthermore, we also tested the sampling attack in other watermarked datasets with different values of the skewness parameter and obtained similar results. 
\subsection{Destroy Attack} \label{destroy}
In this case the attacker $\mathcal{A}$ tries to damage the watermark. 
The no-security-by-obscurity principle~\cite{kerckhoffs1883kerckhoffs} allows $\mathcal{A}$ to know that it can destroy the watermark in a way that it cannot be detected by the owner. 
$\mathcal{A}$ computes the histogram of watermarked data $D_w$. $\mathcal{A}$ modifies the frequencies of \uP s as it pleases by allowing re-ordering (changing the popularity/rank of the \uP s) or without allowing re-ordering. 
We define these two attacks and discuss \textit{FreqyWM}'s robustness  against them. 

\subsubsection{Destroy Attack without re-ordering}\label{destroyworeorder}
In this attack type, $\mathcal{A}$ can modify the frequencies without changing the order of frequencies. 
We introduce two types: (1) attacker changes the frequencies randomly by the given boundaries and (2) attacker changes the frequencies by (at most) some percentage. 

\noindent \textbf{Changing the frequencies randomly within the boundaries.} $\mathcal{A}$ calculates the boundaries for each \uP. Then, $\mathcal{A}$ chooses a random value $r_i$ for each $\uL_i$ as $r_i \leftarrow (-l_i,u_i)$. $\mathcal{A}$ changes the frequency of $\uL_i$ and updates $u_{i+1}$ of $\uL_{i+1}$ by $r_i$.

\noindent \textbf{Changing the frequencies by (at most) some percentage.} $\mathcal{A}$ changes the frequencies of \uP s up to some percentage (e.g., $1\%$). To illustrate, $\mathcal{A}$ calculates the boundaries as $u_i$ and $l_i$ for each $\uL_i$ where it sets the percentage to $1\%$. 
It calculates $u'_i=floor(u_i\times 0.01)$ and $l'_i=floor(l_i \times 0.01)$. 
Then it gets a random value $r_i$ between $(-l'_i,u'_i)$. 
It hereby changes $\uL_i$ by at most $\pm 1\%$.   
After every change ($f'_i=f_i+r_i$), the boundary of the next element is updated. 
Thus, the attack never changes the ranking/ordering since $l'_i$ and $u'_i$ are already in the boundaries.

\begin{figure}[tp]
\centering
%\vspace{-4pt}
\includegraphics[width=0.24\textwidth]{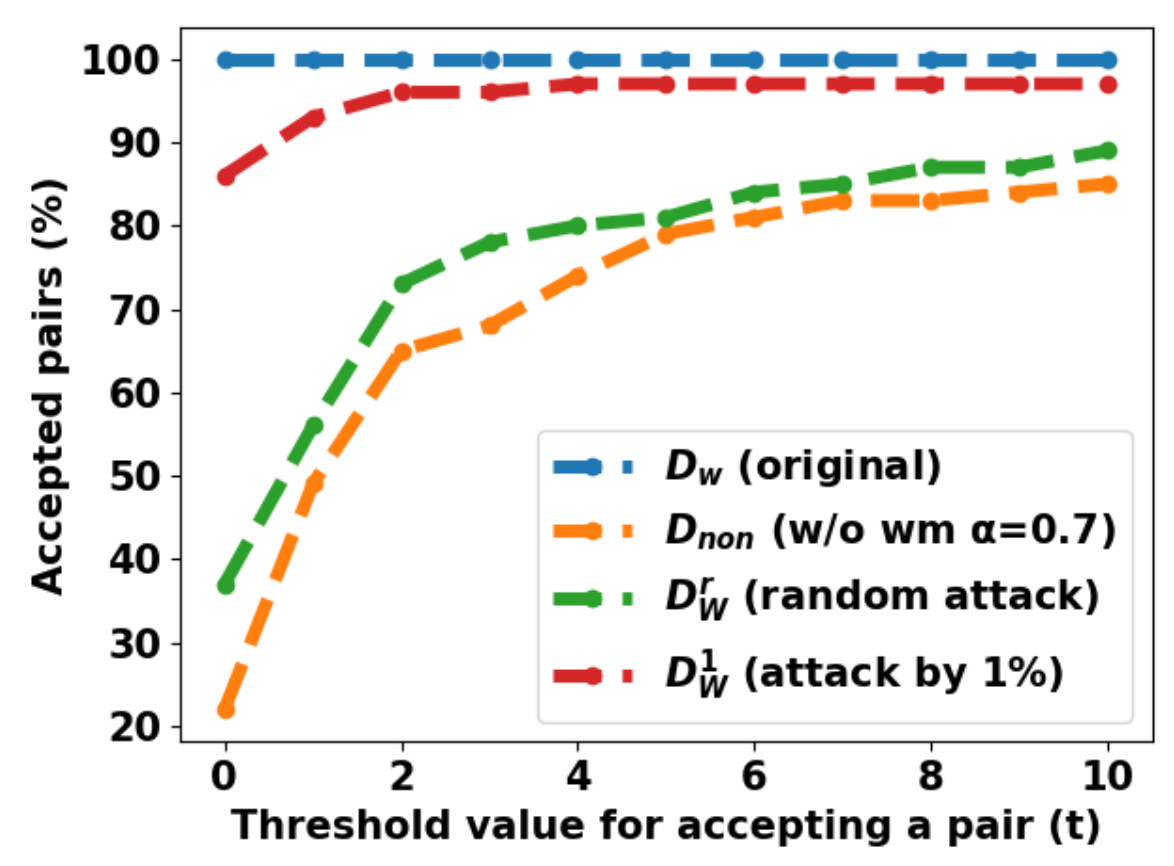}
 \caption{{\small Percentage of verified pairs for the following datasets: (1) $D_w:$ the original watermarked dataset $\alpha=0.5$ without any attack/modification, (2) $D_{non}:$ a non-watermarked dataset defined over the same token space but with $\alpha=0.7$, (3) $D^r_w:$ $D_w$ after attacked by random attack without reordering, (4) $D^1_w:$ $D_w$ after attacked by changing frequencies at most $1\%$. 
 }}
 \label{fig:uncontAttack}
\end{figure}

Figure \ref{fig:uncontAttack} shows how robust \textit{FreqyWM} is against these two destroy attacks. 
We compare the success rate (the percentage of accepted \uP \ pairs given threshold for accepting a pair $t$) of detection algorithm with respect to modified watermarked data after the attacks. 
We also include in the figure a second dataset of skewness $\alpha=0.7$ \emph{that does not carry the watermark}, and report on how many of its pairs would be falsely verified for different values of $t$. 
For an attack in which the frequencies are changed by (at most) some percentage (represented by the red line in the figure), \textit{FreqyWM} can detect around $90\%$ of the pairs when $t=0$. 
When $t$ is increased, after a point where $t \geq f_i-f_j \mod s_{ij}$, the success rate converges at around $90\%$. 
For an attack where the frequencies are changed randomly within the upper and lower frequency boundaries (green line in Figure ~\ref{fig:uncontAttack}), \textit{FreqyWM} can detect more than $35\%$ of the pairs when $t=0$. 
Note that the latter is more powerful than the former attack. 
There is a direct proportion between $t$ and the success rate. 
As shown, the success rate reaches to $90\%$ when $t$ goes to $10$. 

From Figure \ref{fig:uncontAttack}, we can interpret in what parameter setting false negative (rejecting a watermarked pair) and false positive (accepting a pair as watermarked while it is not) can be avoided. 
Thus, the watermarking detection algorithm can successfully detect a watermarked dataset attacked and reject a dataset that was not watermarked. 
For instance, the rate of false positive increases when the threshold for accepting a pair $t$ increases while the minimum number of accepted pairs for detection $k$ decreases which is the area under the results of the dataset (not watermarked) with a different skewness parameter (the area under the orange line).  
On the other hand, the rate of false negative increases when the threshold accepting a pair $t$ decreases while threshold for detecting a watermark $k$ increases which is the area above the results of the attack without re-ordering (the area above the green line) if we consider a very strong attack. 
To avoid false negatives/positives, convenient parameter settings (i.e., $t$ and $k$) for detecting a watermark lie between these two areas (between the orange and the green lines in Figure \ref{fig:uncontAttack}). 
However, if a weaker attack (changing the frequencies by some percentage ) is considered, the range of these parameters increases (the area between the red and orange line). 
Hence, the detection algorithm can detect a watermarked dataset and reject a dataset not watermarked by the owner with a careful parameter setting. 
\revised{For instance, adjusting $t$ (and $k$) based on the nature of the data and the specific application context can enable us to reduce the false positives/negatives.} This is an interesting future work.

\subsubsection{Destroy Attack with re-ordering} 
In this attack type, an attacker $\mathcal{A}$ can modify the frequencies as it pleases without observing any ordering restrictions. Note that this attack introduces more noise than the attack without re-ordering which reduces the usability of watermarked data $D_w$.  
$\mathcal{A}$ modifies the frequencies with various percentages $ [10\%,30\%,50\%,60\%,80\%,90\%]$ where the success rates are $[94\% , 88\% , 82\% , 79\%, 78\% , 76\%]$ respectively. 
\textit{FreqyWM} can detect the watermark with $76\%$ chance up to modifications of $90\%$ in frequencies approximately (where $t=4$).

\subsection{Re-watermarking/ False-Claim Attack}\label{falseclaim} 
This attack is mounted by an attacker $\mathcal{A}$ creating a new watermark on the watermarked data $D_w$, generated by an honest owner. 
$\mathcal{A}$ generates its own watermarked data by simply inserting $D_w$ into the watermark generation algorithm as data to produce $D^\mathcal{A}_w$. 
Then $\mathcal{A}$ can present $D^\mathcal{A}_w$ and claim the ownership of $D^\mathcal{A}_w$ (since $\mathcal{A}$ can prove its ownership claim by introducing its watermarking secret list $L^\mathcal{A}_{sc}$). 
This attack creates a dispute since both the real owner, who created $D_w$, and $\mathcal{A}$ have proofs of their ownerships. 
The dispute can be arbitrated by introducing a judge (a trusted third party as suggested by \cite{adelsbach2003watermarking}) to the watermarking scheme. 
Both parties, $\mathcal{A}$ and the real owner, introduce their secrets and their watermarked data. 
$\mathcal{A}$ sends its secrets $L^\mathcal{A}_{sc}$ and its watermarked data $D^\mathcal{A}_w$. The real owner sends its secrets $L_{sc}$ and its watermarked data $D_w$. 
The judge computes watermark detection algorithm on each received data for each secret which creates four outputs. 
The judge compares these results and identifies the real owner since only the secret of the real owner can produce accept on both data. 
\revised{To show practicality of our defense against the re-watermarking attack, we implemented the attack above. 
Our results show that the first watermark is detected with $92\%$ on $D^\mathcal{A}_w$ under $t=0$.}
The attacker's only way to succeed is to perform successful guess or destroy attack which it cannot perform as shown previously. 

\begin{figure*}[htp]
\centering
\includegraphics[width=\textwidth]{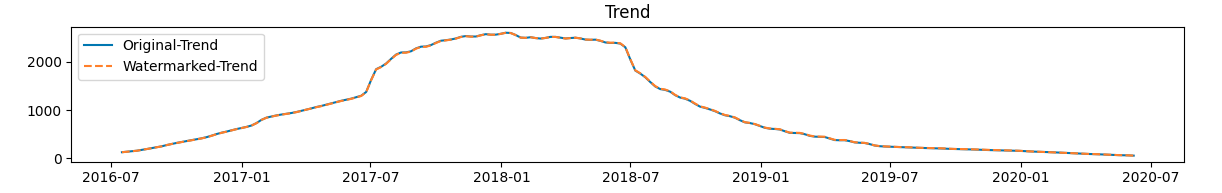}
 \caption{Analysis of the effect of multi-watermarks (i.e., $10$ watermarks) on the trend analysis of \texttt{eyeWnder}. }
 \label{fig:trend}
\end{figure*}

\begin{figure*}[htp]
\centering
\includegraphics[width=\textwidth]{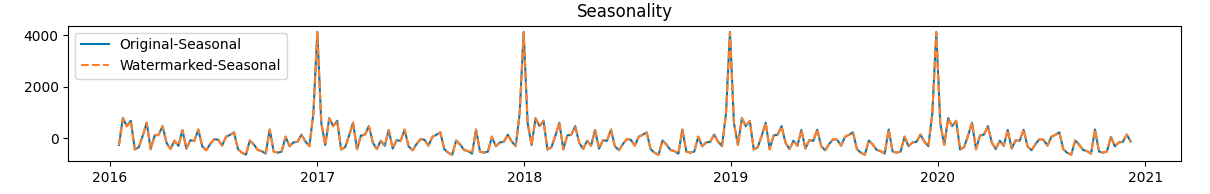}
 \caption{Analysis of the effect of multi-watermarks (i.e., $10$ watermarks) on the seasonality analysis of \texttt{eyeWnder}. }
 \label{fig:seasionality}
\end{figure*}

\begin{figure*}[htp]
\centering
\includegraphics[width=\textwidth]{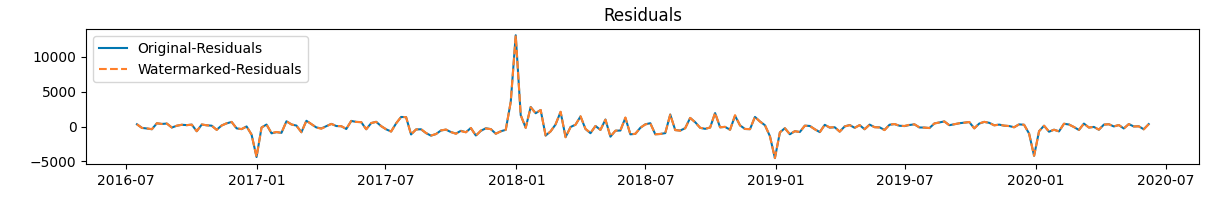}
 \caption{Analysis of the effect of multi-watermarks (i.e., $10$ watermarks) on the residual analysis of \texttt{eyeWnder}. }
 \label{fig:residuals}
\end{figure*}

\begin{figure*}[htp]
\centering
\includegraphics[width=\textwidth]{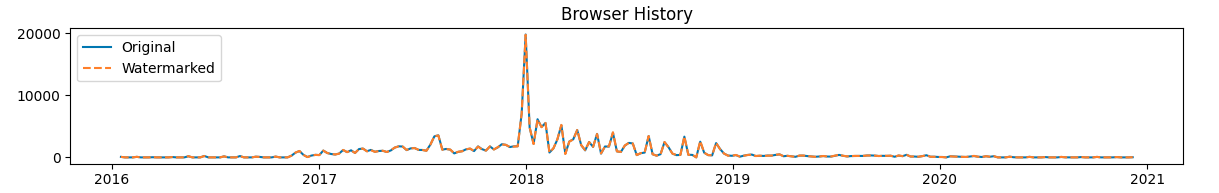}
 \caption{Analysis of the effect of multi-watermarks (i.e., $10$ watermarks) on the browser history analysis of \texttt{eyeWnder}. }
 \label{fig:browser_history}
\end{figure*}
\section{Discussion}\label{discussion}
In this section, we propose possible adjustments to \emph{FreqyWM} for more sophisticated properties and some challenges: \\
\revised{\noindent $\bullet$ \textit{Incremental FreqyWM.} In the literature, there exist watermarking techniques that allow to update a watermark on a dataset without computing insertion from scratch~\cite{chang2021reversible}. 
We believe that an incremental FreqyWM can be built on top of dynamic maximum weighted matching~\cite{Behnezhad23,Solomon16} works but we leave such investigations to future work.}\\
\noindent $\bullet$ \textit{Multi-watermarks.} There are at least two reasons that someone may want to watermark a file multiple times: 1) a legitimate one, e.g., to track the provenance of dataset as it passes from a distributed pipeline, in which a watermark can be added to signify the completion of each one of the processing stages, or to have a chronological order in the versions; 2) a malicious one, to falsely claim ownership via a re-watermarking attack as already discussed in Section~\ref{falseclaim}.
Non-withstanding the motivation, we have run an experiment by running $10$ watermark insertions assuming a budget $b=2$ for each iteration on a sample dataset with skewness $\alpha=0.5$ to calculate three effects: 1) the discrepancy between the original asset and the final one; 2) dataset feature analysis in terms of trend, seasonality, and residuals; and 3) the effect of successive watermarking on machine learning model accuracy.\\
\noindent\textit{Discrepancy.} The resulting similarity between
the original (histogram) and the latest watermarked version is $0.003\%$. 
As it is evident, FreqyWM did not introduce $20\%$ but rather very tiny distortion.
This shows that successive rewatermaking can be practical with \emph{FreqyWM} but we leave further investigation on this matter to future work.\\
\noindent \textit{Feature.} We analyze the change of various features of the \texttt{eyeWnder} dataset that commonly deployed after the watermarking. 
Figure \ref{fig:browser_history} shows the change in the browser history while Figures \ref{fig:trend}, \ref{fig:seasionality}, and \ref{fig:residuals} show the effect of watermarking when the dataset is analyzed in terms of trends, seasonality, and residuals, respectively. 
As shown by the figures, multi-watermarks introduced an insignificant change to the dataset.\\
\noindent \textit{Accuracy.} We analyze the effect of multi-watermarks on an ML model accuracy using the \texttt{eyeWnder} dataset.
We implement a sequential data analysis approach using TensorFlow to predict the next URL in a sequence, utilizing a dataset of timestamped URLs. 
The model consists of an embedding layer, LSTM layers, and a sigmoid output layer. 
We train the model with $10$ epochs and with $128$ batch-size.
The model achieved an accuracy of $82.33\%$ when trained on the original dataset whilst achieving an accuracy of $82.34\%$ when it is watermarked. 
The model trained on the watermarked dataset has slightly better accuracy. 
We suspect that this is due to increase in the size of the dataset (i.e., the watermarked one has $140$ more URLs). 
While our initial results are promising, we plan to extensively investigate this effect using different ML models for more concrete reductions.\\
\noindent $\bullet$ \textit{Challenging datasets.} Apart from datasets with close to uniform frequencies, \emph{FreqyWM} can also be challenged when the range of token values is too wide, e.g., sales' datasets with many decimal values, resulting to very few (if any) repetition of the same value. One natural solution to this is to first bucketize (cluster) the widely ranged data and then apply \emph{FreqyWM} at the level of the bucket as opposed to the exact token value.
\section{Conclusions and Future Work}\label{further}
We proposed \emph{FreqyWM}, a novel frequency watermarking technique for protecting the ownership of data in the emerging new data economy. 
We analysed the performance of \emph{FreqyWM} and showed how \emph{FreqyWM} can encode watermarks with minimal distortion on the original data, provided that the data has sufficient variability in terms of token frequencies.  
We analysed \emph{FreqyWM}'s robustness to generic attacks. 
\revised{\emph{FreqyWM} is applicable to large numbers of tuples sold in wholesale manner in modern DMs. An interesting, yet challenging, research direction is to consider how to watermark small sets or even individual tuples used in distributed data operations such as replication and remote hosting and/or query execution.} 
We are currently looking at more attack scenarios and at devising systematic procedures for optimizing the parameters \revised{and also how to apply \emph{FreqyWM} to multidimensional datasets by overcoming the challenges mentioned in Section \ref{applicability}}. We also investigate integrating data privacy (e.g., differentially-private fingerprinting \cite{ji2022differentially}).
\section*{Acknowledgment}
D.İ. was supported by the European Union’s HORIZON project DataBri-X (101070069). N.L. was supported by the MLEDGE project (REGAGE22e00052829516), funded by the Ministry of Economic Affairs and Digital Transformation and the European Union-NextGenerationEU/PRTR.


\begin{thebibliography}{70}
\providecommand{\natexlab}[1]{#1}
\providecommand{\url}[1]{#1}
\csname url@samestyle\endcsname
\providecommand{\newblock}{\relax}
\providecommand{\bibinfo}[2]{#2}
\providecommand{\BIBentrySTDinterwordspacing}{\spaceskip=0pt\relax}
\providecommand{\BIBentryALTinterwordstretchfactor}{4}
\providecommand{\BIBentryALTinterwordspacing}{\spaceskip=\fontdimen2\font plus
\BIBentryALTinterwordstretchfactor\fontdimen3\font minus \fontdimen4\font\relax}
\providecommand{\BIBforeignlanguage}[2]{{%
\expandafter\ifx\csname l@#1\endcsname\relax
\typeout{** WARNING: IEEEtranN.bst: No hyphenation pattern has been}%
\typeout{** loaded for the language `#1'. Using the pattern for}%
\typeout{** the default language instead.}%
\else
\language=\csname l@#1\endcsname
\fi
#2}}
\providecommand{\BIBdecl}{\relax}
\BIBdecl

\bibitem[Azcoitia and Laoutaris(2022)]{azcoitia2022survey}
\BIBentryALTinterwordspacing
S.~A. Azcoitia and N.~Laoutaris, ``A survey of data marketplaces and their business models,'' \emph{{SIGMOD} Rec.}, vol.~51, no.~3, pp. 18--29, 2022. [Online]. Available: \url{https://doi.org/10.1145/3572751.3572755}
\BIBentrySTDinterwordspacing

\bibitem[Lutu et~al.(2020)Lutu, Perino, Bagnulo, Frias-Martinez, and Khangosstar]{10.1145/3419394.3423655}
\BIBentryALTinterwordspacing
A.~Lutu, D.~Perino, M.~Bagnulo, E.~Frias-Martinez, and J.~Khangosstar, ``A characterization of the covid-19 pandemic impact on a mobile network operator traffic,'' in \emph{Proceedings of the ACM Internet Measurement Conference}, ser. IMC '20.\hskip 1em plus 0.5em minus 0.4em\relax New York, NY, USA: Association for Computing Machinery, 2020. [Online]. Available: \url{https://doi.org/10.1145/3419394.3423655}
\BIBentrySTDinterwordspacing

\bibitem[Evans et~al.(2018)Evans, Kolesnikov, and Rosulek]{mpcevans18}
\BIBentryALTinterwordspacing
D.~Evans, V.~Kolesnikov, and M.~Rosulek, ``A pragmatic introduction to secure multi-party computation,'' \emph{Found. Trends Priv. Secur.}, 2018. [Online]. Available: \url{https://doi.org/10.1561/3300000019}
\BIBentrySTDinterwordspacing

\bibitem[Gentry(2009)]{fheGentry09}
\BIBentryALTinterwordspacing
C.~Gentry, ``A fully homomorphic encryption scheme,'' Ph.D. dissertation, Stanford University, {USA}, 2009. [Online]. Available: \url{https://searchworks.stanford.edu/view/8493082}
\BIBentrySTDinterwordspacing

\bibitem[Boneh et~al.(2011)Boneh, Sahai, and Waters]{bonehfe11}
\BIBentryALTinterwordspacing
D.~Boneh, A.~Sahai, and B.~Waters, ``Functional encryption: Definitions and challenges,'' in \emph{Theory of Cryptography Conference, {TCC}}.\hskip 1em plus 0.5em minus 0.4em\relax Springer, 2011. [Online]. Available: \url{https://doi.org/10.1007/978-3-642-19571-6\_16}
\BIBentrySTDinterwordspacing

\bibitem[Sabt et~al.(2015)Sabt, Achemlal, and Bouabdallah]{teewhatnot15}
\BIBentryALTinterwordspacing
M.~Sabt, M.~Achemlal, and A.~Bouabdallah, ``Trusted execution environment: What it is, and what it is not,'' in \emph{TrustCom/BigDataSE/ISPA}.\hskip 1em plus 0.5em minus 0.4em\relax {IEEE}, 2015. [Online]. Available: \url{https://doi.org/10.1109/Trustcom.2015.357}
\BIBentrySTDinterwordspacing

\bibitem[Li et~al.(2021)Li, Ghosh, Gupta, Mehrotra, Panwar, and Sharma]{prism2021}
\BIBentryALTinterwordspacing
Y.~Li, D.~Ghosh, P.~Gupta, S.~Mehrotra, N.~Panwar, and S.~Sharma, ``{PRISM:} private verifiable set computation over multi-owner outsourced databases,'' in \emph{SIGMOD: International Conference on Management of Data, Virtual}.\hskip 1em plus 0.5em minus 0.4em\relax {ACM}, 2021. [Online]. Available: \url{https://doi.org/10.1145/3448016.3452839}
\BIBentrySTDinterwordspacing

\bibitem[Poddar et~al.(2019)Poddar, Boelter, and Popa]{arxvldb2019}
\BIBentryALTinterwordspacing
R.~Poddar, T.~Boelter, and R.~A. Popa, ``Arx: An encrypted database using semantically secure encryption,'' \emph{Proc. {VLDB} Endow.}, 2019. [Online]. Available: \url{http://www.vldb.org/pvldb/vol12/p1664-poddar.pdf}
\BIBentrySTDinterwordspacing

\bibitem[Anciaux et~al.(2019)Anciaux, Bouganim, Pucheral, Popa, and Scerri]{teeAnciaux19}
\BIBentryALTinterwordspacing
N.~Anciaux, L.~Bouganim, P.~Pucheral, I.~S. Popa, and G.~Scerri, ``Personal database security and trusted execution environments: {A} tutorial at the crossroads,'' \emph{Proc. {VLDB} Endow.}, 2019. [Online]. Available: \url{http://www.vldb.org/pvldb/vol12/p1994-anciaux.pdf}
\BIBentrySTDinterwordspacing

\bibitem[Ren et~al.(2022)Ren, Su, Gu, Wang, Li, Xie, Bian, Li, and Zhang]{heda2022}
\BIBentryALTinterwordspacing
X.~Ren, L.~Su, Z.~Gu, S.~Wang, F.~Li, Y.~Xie, S.~Bian, C.~Li, and F.~Zhang, ``{HEDA:} multi-attribute unbounded aggregation over homomorphically encrypted database,'' \emph{Proc. {VLDB} Endow.}, 2022. [Online]. Available: \url{https://www.vldb.org/pvldb/vol16/p601-gu.pdf}
\BIBentrySTDinterwordspacing

\bibitem[Zhou et~al.(2021{\natexlab{a}})Zhou, Cai, Peng, Wang, Ma, and Li]{veridb21}
\BIBentryALTinterwordspacing
W.~Zhou, Y.~Cai, Y.~Peng, S.~Wang, K.~Ma, and F.~Li, ``Veridb: An sgx-based verifiable database,'' in \emph{{SIGMOD}: International Conference on Management of Data}.\hskip 1em plus 0.5em minus 0.4em\relax {ACM}, 2021. [Online]. Available: \url{https://doi.org/10.1145/3448016.3457308}
\BIBentrySTDinterwordspacing

\bibitem[Jougleux(2022)]{dataownershiplaw2022}
P.~Jougleux, ``Data ownership (and succession law),'' in \emph{Facebook and the (EU) Law: How the Social Network Reshaped the Legal Framework}.\hskip 1em plus 0.5em minus 0.4em\relax Springer, 2022, pp. 129--143.

\bibitem[Kennedy et~al.(2022)Kennedy, Subramaniam, Galhotra, and Fernandez]{dmrevisit22}
\BIBentryALTinterwordspacing
J.~Kennedy, P.~Subramaniam, S.~Galhotra, and R.~C. Fernandez, ``Revisiting online data markets in 2022: {A} seller and buyer perspective,'' \emph{{SIGMOD} Rec.}, vol.~51, no.~3, pp. 30--37, 2022. [Online]. Available: \url{https://doi.org/10.1145/3572751.3572757}
\BIBentrySTDinterwordspacing

\bibitem[Fernandez et~al.(2020)Fernandez, Subramaniam, and Franklin]{dmassets20}
\BIBentryALTinterwordspacing
R.~C. Fernandez, P.~Subramaniam, and M.~J. Franklin, ``Data market platforms: Trading data assets to solve data problems,'' \emph{Proc. {VLDB} Endow.}, vol.~13, no.~11, pp. 1933--1947, 2020. [Online]. Available: \url{http://www.vldb.org/pvldb/vol13/p1933-fernandez.pdf}
\BIBentrySTDinterwordspacing

\bibitem[Banterle(2020)]{banterle2020data}
\BIBentryALTinterwordspacing
F.~Banterle, ``Data ownership in the data economy: a european dilemma,'' \emph{EU Internet Law in the Digital Era: Regulation and Enforcement}, pp. 199--225, 2020. [Online]. Available: \url{https://papers.ssrn.com/sol3/papers.cfm?abstract_id=3277330}
\BIBentrySTDinterwordspacing

\bibitem[Asikuzzaman and Pickering(2018)]{asikuzzaman2017overview}
\BIBentryALTinterwordspacing
M.~Asikuzzaman and M.~R. Pickering, ``An overview of digital video watermarking,'' \emph{{IEEE} Trans. Circuits Syst. Video Technol.}, 2018. [Online]. Available: \url{https://doi.org/10.1109/TCSVT.2017.2712162}
\BIBentrySTDinterwordspacing

\bibitem[Begum and Uddin(2020)]{begum2020digital}
\BIBentryALTinterwordspacing
M.~Begum and M.~S. Uddin, ``Digital image watermarking techniques: {A} review,'' \emph{Inf.}, 2020. [Online]. Available: \url{https://doi.org/10.3390/info11020110}
\BIBentrySTDinterwordspacing

\bibitem[Ma et~al.(2019)Ma, Jia, Li, Zheng, and Wu]{softwarewmtifs}
\BIBentryALTinterwordspacing
H.~Ma, C.~Jia, S.~Li, W.~Zheng, and D.~Wu, ``Xmark: Dynamic software watermarking using collatz conjecture,'' \emph{{IEEE} Trans. Inf. Forensics Secur.}, 2019. [Online]. Available: \url{https://doi.org/10.1109/TIFS.2019.2908071}
\BIBentrySTDinterwordspacing

\bibitem[Zhou et~al.(2005)Zhou, Pang, Tan, and Mangla]{xmlwatermarking}
\BIBentryALTinterwordspacing
X.~Zhou, H.~Pang, K.~Tan, and D.~Mangla, ``Wmxml: {A} system for watermarking {XML} data,'' in \emph{International Conference on Very Large Data Bases (VLDB)}.\hskip 1em plus 0.5em minus 0.4em\relax {ACM}, 2005. [Online]. Available: \url{http://www.vldb.org/conf/2005/papers/p1318-zhou.pdf}
\BIBentrySTDinterwordspacing

\bibitem[Agrawal and Kiernan(2002)]{agrawal2002watermarking}
\BIBentryALTinterwordspacing
R.~Agrawal and J.~Kiernan, ``Watermarking relational databases,'' in \emph{Proceedings ofInternational Conference on Very Large Data Bases, {VLDB}}, 2002. [Online]. Available: \url{http://www.vldb.org/conf/2002/S05P03.pdf}
\BIBentrySTDinterwordspacing

\bibitem[Agrawal et~al.(2003{\natexlab{a}})Agrawal, Haas, and Kiernan]{agrawalsigmod}
\BIBentryALTinterwordspacing
R.~Agrawal, P.~J. Haas, and J.~Kiernan, ``A system for watermarking relational databases,'' in \emph{{ACM} {SIGMOD} International Conference}, 2003. [Online]. Available: \url{https://doi.org/10.1145/872757.872865}
\BIBentrySTDinterwordspacing

\bibitem[Wang and Kerschbaum(2021)]{wang2021riga}
\BIBentryALTinterwordspacing
T.~Wang and F.~Kerschbaum, ``{RIGA:} covert and robust white-box watermarking of deep neural networks,'' in \emph{{WWW}: The Web Conference}, 2021. [Online]. Available: \url{https://doi.org/10.1145/3442381.3450000}
\BIBentrySTDinterwordspacing

\bibitem[Rani and Halder(2022)]{RaniH22}
\BIBentryALTinterwordspacing
S.~Rani and R.~Halder, ``Comparative analysis of relational database watermarking techniques: An empirical study,'' \emph{{IEEE} Access}, vol.~10, pp. 27\,970--27\,989, 2022. [Online]. Available: \url{https://doi.org/10.1109/ACCESS.2022.3157866}
\BIBentrySTDinterwordspacing

\bibitem[Agarwal et~al.(2019)Agarwal, Singh, and Singh]{agarwal2019survey}
\BIBentryALTinterwordspacing
N.~Agarwal, A.~K. Singh, and P.~K. Singh, ``Survey of robust and imperceptible watermarking,'' \emph{Multim. Tools Appl.}, 2019. [Online]. Available: \url{https://doi.org/10.1007/s11042-018-7128-5}
\BIBentrySTDinterwordspacing

\bibitem[Agrawal et~al.(2003{\natexlab{b}})Agrawal, Haas, and Kiernan]{agrawal2003watermarking}
\BIBentryALTinterwordspacing
R.~Agrawal, P.~J. Haas, and J.~Kiernan, ``Watermarking relational data: framework, algorithms and analysis,'' \emph{{VLDB} J.}, 2003. [Online]. Available: \url{https://doi.org/10.1007/s00778-003-0097-x}
\BIBentrySTDinterwordspacing

\bibitem[Ji et~al.(2021{\natexlab{a}})Ji, Yilmaz, Ayday, and Li]{ji2021curse}
\BIBentryALTinterwordspacing
T.~Ji, E.~Yilmaz, E.~Ayday, and P.~Li, ``The curse of correlations for robust fingerprinting of relational databases,'' in \emph{{RAID} : International Symposium on Research in Attacks, Intrusions and Defenses}.\hskip 1em plus 0.5em minus 0.4em\relax {ACM}, 2021. [Online]. Available: \url{https://doi.org/10.1145/3471621.3471853}
\BIBentrySTDinterwordspacing

\bibitem[Quiring et~al.(2018)Quiring, Arp, and Rieck]{quiring2018forgotten}
\BIBentryALTinterwordspacing
E.~Quiring, D.~Arp, and K.~Rieck, ``Forgotten siblings: Unifying attacks on machine learning and digital watermarking,'' in \emph{{IEEE} European Symposium on Security and Privacy, EuroS{\&}P}.\hskip 1em plus 0.5em minus 0.4em\relax {IEEE}, 2018. [Online]. Available: \url{https://doi.org/10.1109/EuroSP.2018.00041}
\BIBentrySTDinterwordspacing

\bibitem[Cohen et~al.(2018)Cohen, Holmgren, Nishimaki, Vaikuntanathan, and Wichs]{cohen2018watermarking}
\BIBentryALTinterwordspacing
A.~Cohen, J.~Holmgren, R.~Nishimaki, V.~Vaikuntanathan, and D.~Wichs, ``Watermarking cryptographic capabilities,'' \emph{{SIAM} J. Comput.}, 2018. [Online]. Available: \url{https://doi.org/10.1137/18M1164834}
\BIBentrySTDinterwordspacing

\bibitem[Tang et~al.(2021)Tang, Cao, Dong, and Shen]{bigdb21}
\BIBentryALTinterwordspacing
X.~Tang, Z.~Cao, X.~Dong, and J.~Shen, ``Pkmark: {A} robust zero-distortion blind reversible scheme for watermarking relational databases,'' in \emph{{IEEE} International Conference on Big Data Science and Engineering}, 2021. [Online]. Available: \url{https://doi.org/10.1109/BigDataSE53435.2021.00020}
\BIBentrySTDinterwordspacing

\bibitem[Li et~al.(2022)Li, Li, Yan, Zhang, Yu, and Long]{dbwmli22}
W.~Li, N.~Li, J.~Yan, Z.~Zhang, P.~Yu, and G.~Long, ``Secure and high-quality watermarking algorithms for relational database based on semantic,'' \emph{IEEE Transactions on Knowledge and Data Engineering}, pp. 1--14, 2022.

\bibitem[Gort et~al.(2021)Gort, Olliaro, Cortesi, and Uribe]{textwm2021}
\BIBentryALTinterwordspacing
M.~L.~P. Gort, M.~Olliaro, A.~Cortesi, and C.~F. Uribe, ``Semantic-driven watermarking of relational textual databases,'' \emph{Expert Syst. Appl.}, 2021. [Online]. Available: \url{https://doi.org/10.1016/j.eswa.2020.114013}
\BIBentrySTDinterwordspacing

\bibitem[Lin et~al.(2021)Lin, Nguyen, and Chang]{categorywm21}
\BIBentryALTinterwordspacing
C.~Lin, T.~Nguyen, and C.~Chang, ``{LRW-CRDB:} lossless robust watermarking scheme for categorical relational databases,'' \emph{Symmetry}, 2021. [Online]. Available: \url{https://doi.org/10.3390/sym13112191}
\BIBentrySTDinterwordspacing

\bibitem[Kumar et~al.(2020)Kumar, Singh, and Yadav]{kumar2020recent}
\BIBentryALTinterwordspacing
S.~Kumar, B.~K. Singh, and M.~Yadav, ``A recent survey on multimedia and database watermarking,'' \emph{Multim. Tools Appl.}, vol.~79, no. 27-28, pp. 20\,149--20\,197, 2020. [Online]. Available: \url{https://doi.org/10.1007/s11042-020-08881-y}
\BIBentrySTDinterwordspacing

\bibitem[Jony et~al.(2021)Jony, Johora, and Katha]{jony2021robust}
\BIBentryALTinterwordspacing
M.~H. Jony, F.~T. Johora, and J.~F. Katha, ``A robust and efficient numeric approach for relational database watermarking,'' in \emph{{IEEE} International Conference on Sustainable Technologies for Industry 4.0 (STI)}, 2021. [Online]. Available: \url{https://ieeexplore.ieee.org/document/9732582}
\BIBentrySTDinterwordspacing

\bibitem[Shehab et~al.(2008)Shehab, Bertino, and Ghafoor]{shehab2007watermarking}
\BIBentryALTinterwordspacing
M.~Shehab, E.~Bertino, and A.~Ghafoor, ``Watermarking relational databases using optimization-based techniques,'' \emph{{IEEE} Trans. Knowl. Data Eng.}, 2008. [Online]. Available: \url{https://doi.org/10.1109/TKDE.2007.190668}
\BIBentrySTDinterwordspacing

\bibitem[Ibosiola et~al.(2018)Ibosiola, Steer, Garc{\'{\i}}a{-}Recuero, Stringhini, Uhlig, and Tyson]{ibosiola2018movie}
\BIBentryALTinterwordspacing
D.~Ibosiola, B.~A. Steer, {\'{A}}.~Garc{\'{\i}}a{-}Recuero, G.~Stringhini, S.~Uhlig, and G.~Tyson, ``Movie pirates of the caribbean: Exploring illegal streaming cyberlockers,'' in \emph{Proceedings of the Twelfth International Conference on Web and Social Media, {ICWSM}}.\hskip 1em plus 0.5em minus 0.4em\relax {AAAI} Press, 2018. [Online]. Available: \url{https://aaai.org/ocs/index.php/ICWSM/ICWSM18/paper/view/17835}
\BIBentrySTDinterwordspacing

\bibitem[Zhou et~al.(2021{\natexlab{b}})Zhou, Hu, and Wang]{8003392}
\BIBentryALTinterwordspacing
W.~Zhou, J.~Hu, and S.~Wang, ``Enhanced locality-sensitive hashing for fingerprint forensics over large multi-sensor databases,'' \emph{{IEEE} Trans. Big Data}, 2021. [Online]. Available: \url{https://doi.org/10.1109/TBDATA.2017.2736547}
\BIBentrySTDinterwordspacing

\bibitem[Lei et~al.(2020)Lei, Huang, Kankanhalli, and Tung]{localhashing20}
\BIBentryALTinterwordspacing
Y.~Lei, Q.~Huang, M.~S. Kankanhalli, and A.~K.~H. Tung, ``Locality-sensitive hashing scheme based on longest circular co-substring,'' in \emph{Proceedings of the 2020 International Conference on Management of Data, {SIGMOD}}.\hskip 1em plus 0.5em minus 0.4em\relax {ACM}, 2020. [Online]. Available: \url{https://doi.org/10.1145/3318464.3389778}
\BIBentrySTDinterwordspacing

\bibitem[Chang et~al.(2019)Chang, Ghosh, Sanadhya, Singh, and White]{chang2019fbhash}
\BIBentryALTinterwordspacing
D.~Chang, M.~Ghosh, S.~K. Sanadhya, M.~Singh, and D.~R. White, ``Fbhash: {A} new similarity hashing scheme for digital forensics,'' \emph{Digit. Investig.}, 2019. [Online]. Available: \url{https://doi.org/10.1016/j.diin.2019.04.006}
\BIBentrySTDinterwordspacing

\bibitem[Osiakwan and Akl(1995)]{osiakwan1990maximum}
\BIBentryALTinterwordspacing
C.~N.~K. Osiakwan and S.~G. Akl, ``The maximum weight perfect matching problem for complete weighted graphs is in pc*,'' \emph{Parallel Algorithms Appl.}, 1995. [Online]. Available: \url{https://doi.org/10.1080/10637199508915506}
\BIBentrySTDinterwordspacing

\bibitem[Galil(1986)]{galil1986maxwm}
Z.~Galil, ``Efficient algorithms for finding maximum matching in graphs,'' in \emph{ACM CSUR}, 1986.

\bibitem[Cormen et~al.(2009)Cormen, Leiserson, Rivest, and Stein]{cormen2009introduction}
T.~H. Cormen, C.~E. Leiserson, R.~L. Rivest, and C.~Stein, \emph{Introduction to algorithms}.\hskip 1em plus 0.5em minus 0.4em\relax MIT press, 2009.

\bibitem[Ayday et~al.(2019)Ayday, Yilmaz, and Yilmaz]{ayday2019robust}
\BIBentryALTinterwordspacing
E.~Ayday, E.~Yilmaz, and A.~Yilmaz, ``Robust optimization-based watermarking scheme for sequential data,'' in \emph{International Symposium on Research in Attacks, Intrusions and Defenses, {RAID}}, 2019. [Online]. Available: \url{https://www.usenix.org/conference/raid2019/presentation/ayday}
\BIBentrySTDinterwordspacing

\bibitem[Ji et~al.(2022)Ji, Ayday, Yilmaz, and Li]{ji2022robust}
\BIBentryALTinterwordspacing
T.~Ji, E.~Ayday, E.~Yilmaz, and P.~Li, ``Robust fingerprinting of genomic databases,'' \emph{CoRR}, vol. abs/2204.01801, 2022. [Online]. Available: \url{https://doi.org/10.48550/arXiv.2204.01801}
\BIBentrySTDinterwordspacing

\bibitem[Kamran and Farooq(2018)]{kamran2018comprehensive}
M.~Kamran and M.~Farooq, ``A comprehensive survey of watermarking relational databases research,'' in \emph{arXiv preprint arXiv:1801.08271}, 2018.

\bibitem[Panah et~al.(2016)Panah, van Schyndel, Sellis, and Bertino]{panah2016properties}
\BIBentryALTinterwordspacing
A.~S. Panah, R.~G. van Schyndel, T.~K. Sellis, and E.~Bertino, ``On the properties of non-media digital watermarking: {A} review of state of the art techniques,'' \emph{{IEEE} Access}, 2016. [Online]. Available: \url{https://doi.org/10.1109/ACCESS.2016.2570812}
\BIBentrySTDinterwordspacing

\bibitem[Farfoura et~al.(2012)Farfoura, Horng, Lai, Run, Chen, and Khan]{farfoura2012blind}
\BIBentryALTinterwordspacing
M.~E. Farfoura, S.~Horng, J.~Lai, R.~Run, R.~Chen, and M.~K. Khan, ``A blind reversible method for watermarking relational databases based on a time-stamping protocol,'' \emph{Expert Syst. Appl.}, 2012. [Online]. Available: \url{https://doi.org/10.1016/j.eswa.2011.09.005}
\BIBentrySTDinterwordspacing

\bibitem[Li and Deng(2006)]{li2006publicly}
\BIBentryALTinterwordspacing
Y.~Li and R.~H. Deng, ``Publicly verifiable ownership protection for relational databases,'' in \emph{Proceedings of th {ACM} Symposium on Information, Computer and Communications Security, {ASIACCS}}.\hskip 1em plus 0.5em minus 0.4em\relax {ACM}, 2006. [Online]. Available: \url{https://doi.org/10.1145/1128817.1128832}
\BIBentrySTDinterwordspacing

\bibitem[Hu et~al.(2019)Hu, Zhao, and Zheng]{hu2018new}
\BIBentryALTinterwordspacing
D.~Hu, D.~Zhao, and S.~Zheng, ``A new robust approach for reversible database watermarking with distortion control,'' \emph{{IEEE} Trans. Knowl. Data Eng.}, 2019. [Online]. Available: \url{https://doi.org/10.1109/TKDE.2018.2851517}
\BIBentrySTDinterwordspacing

\bibitem[El{-}Bakry and Hamada(2010)]{el2010novel}
\BIBentryALTinterwordspacing
H.~M. El{-}Bakry and M.~Hamada, ``A novel watermark technique for relational databases,'' in \emph{Artificial Intelligence and Computational Intelligence - International Conference, {AICI} 2010, Sanya, China, October 23-24, 2010, Proceedings, Part {II}}, ser. Lecture Notes in Computer Science.\hskip 1em plus 0.5em minus 0.4em\relax Springer, 2010. [Online]. Available: \url{https://doi.org/10.1007/978-3-642-16527-6\_29}
\BIBentrySTDinterwordspacing

\bibitem[Darwish et~al.(2018)Darwish, Selim, and El{-}Sherbiny]{darwishdistortion}
\BIBentryALTinterwordspacing
S.~M. Darwish, H.~A. Selim, and M.~M. El{-}Sherbiny, ``Distortion free database watermarking system based on intelligent mechanism for content integrity and ownership control,'' \emph{J. Comput.}, 2018. [Online]. Available: \url{https://doi.org/10.17706/jcp.13.9.1053-1066}
\BIBentrySTDinterwordspacing

\bibitem[Zhang et~al.(2008)Zhang, Yang, and Niu]{zhang2008reversible}
Y.~Zhang, B.~Yang, and X.-M. Niu, ``Reversible watermarking for relational database authentication,'' 2008.

\bibitem[Wang et~al.(2022)Wang, Liu, Wang, and Liang]{wang2022fbipt}
\BIBentryALTinterwordspacing
W.~Wang, C.~Liu, Z.~Wang, and T.~Liang, ``{FBIPT:} {A} new robust reversible database watermarking technique based on position tuples,'' in \emph{International Conference on Data Intelligence and Security, {ICDIS}}.\hskip 1em plus 0.5em minus 0.4em\relax {IEEE}, 2022, pp. 67--74. [Online]. Available: \url{https://doi.org/10.1109/ICDIS55630.2022.00018}
\BIBentrySTDinterwordspacing

\bibitem[Gupta and Pieprzyk(2009)]{gupta2009reversible}
\BIBentryALTinterwordspacing
G.~Gupta and J.~Pieprzyk, ``Reversible and blind database watermarking using difference expansion,'' \emph{Int. J. Digit. Crime Forensics}, 2009. [Online]. Available: \url{https://doi.org/10.4018/jdcf.2009040104}
\BIBentrySTDinterwordspacing

\bibitem[Jawad and Khan(2013)]{jawad2013genetic}
\BIBentryALTinterwordspacing
K.~Jawad and A.~Khan, ``Genetic algorithm and difference expansion based reversible watermarking for relational databases,'' \emph{J. Syst. Softw.}, 2013. [Online]. Available: \url{https://doi.org/10.1016/j.jss.2013.06.023}
\BIBentrySTDinterwordspacing

\bibitem[Imamoglu et~al.(2017)Imamoglu, Ulutas, and Ulutas]{imamoglu2017new}
\BIBentryALTinterwordspacing
M.~B. Imamoglu, M.~Ulutas, and G.~Ulutas, ``A new reversible database watermarking approach with firefly optimization algorithm,'' \emph{Mathematical Problems in Engineering}, 2017. [Online]. Available: \url{https://doi.org/10.1155/2017/1387375}
\BIBentrySTDinterwordspacing

\bibitem[Chang et~al.(2021)Chang, Nguyen, and Lin]{chang2021reversible}
\BIBentryALTinterwordspacing
C.~Chang, T.~Nguyen, and C.~Lin, ``A reversible database watermark scheme for textual and numerical datasets,'' in \emph{{IEEE/ACIS} International Conference on Software Engineering, Artificial Intelligence, Networking and Parallel/Distributed Computing, {SNPD}}.\hskip 1em plus 0.5em minus 0.4em\relax {IEEE}, 2021. [Online]. Available: \url{https://doi.org/10.1109/SNPD51163.2021.9704991}
\BIBentrySTDinterwordspacing

\bibitem[Iordanou et~al.(2019)Iordanou, Kourtellis, Carrascosa, Soriente, Cuevas, and Laoutaris]{iordanou2019beyond}
\BIBentryALTinterwordspacing
C.~Iordanou, N.~Kourtellis, J.~M. Carrascosa, C.~Soriente, R.~Cuevas, and N.~Laoutaris, ``Beyond content analysis: detecting targeted ads via distributed counting,'' in \emph{Proceedings of the 15th International Conference on Emerging Networking Experiments And Technologies, CoNEXT}.\hskip 1em plus 0.5em minus 0.4em\relax {ACM}, 2019. [Online]. Available: \url{https://doi.org/10.1145/3359989.3365428}
\BIBentrySTDinterwordspacing

\bibitem[Cormode et~al.(2021)Cormode, Maddock, and Maple]{frequencyvldb21}
\BIBentryALTinterwordspacing
G.~Cormode, S.~Maddock, and C.~Maple, ``Frequency estimation under local differential privacy,'' \emph{Proc. {VLDB} Endow.}, 2021. [Online]. Available: \url{http://www.vldb.org/pvldb/vol14/p2046-cormode.pdf}
\BIBentrySTDinterwordspacing

\bibitem[Katz and Lindell(2014)]{KatzLindell2014}
\BIBentryALTinterwordspacing
J.~Katz and Y.~Lindell, \emph{Introduction to Modern Cryptography, Second Edition}.\hskip 1em plus 0.5em minus 0.4em\relax {CRC} Press, 2014. [Online]. Available: \url{https://www.crcpress.com/Introduction-to-Modern-Cryptography-Second-Edition/Katz-Lindell/p/book/9781466570269}
\BIBentrySTDinterwordspacing

\bibitem[{\.I}{\c{s}}ler et~al.(2022){\.I}{\c{s}}ler, Cabana, Garcia-Recuero, Koutrika, and Laoutaris]{icsler2022freqywm}
D.~{\.I}{\c{s}}ler, E.~Cabana, A.~Garcia-Recuero, G.~Koutrika, and N.~Laoutaris, ``Freqywm: Frequency watermarking for the new data economy,'' IMDEA Networks Technical Report, Tech. Rep., 2022.

\bibitem[chi(2022)]{chicagotaxi}
``{Chicago Data Portal},'' 2022, \url{https://data.cityofchicago.org/Transportation/Taxi-Trips/wrvz-psew}.

\bibitem[adu(1996)]{adult}
``{Adult Dataset},'' 1996, \url{https://archive.ics.uci.edu/ml/datasets/Adult}.

\bibitem[Clauset et~al.(2009)Clauset, Shalizi, and Newman]{clauset2009power}
\BIBentryALTinterwordspacing
A.~Clauset, C.~R. Shalizi, and M.~E.~J. Newman, ``Power-law distributions in empirical data,'' \emph{{SIAM} Rev.}, 2009. [Online]. Available: \url{https://doi.org/10.1137/070710111}
\BIBentrySTDinterwordspacing

\bibitem[Goldberg and Sastry(2007)]{goldberg2007genetic}
D.~Goldberg and K.~Sastry, \emph{Genetic algorithms: the design of innovation}.\hskip 1em plus 0.5em minus 0.4em\relax Springer, 2007.

\bibitem[Kerckhoffs(1883)]{kerckhoffs1883kerckhoffs}
A.~Kerckhoffs, ``A. kerckhoffs, la cryptographie militaire, journal des sciences militaires ix, 38 (1883),'' in \emph{Journal des sciences militaires}, 1883.

\bibitem[Adelsbach et~al.(2003)Adelsbach, Katzenbeisser, and Veith]{adelsbach2003watermarking}
\BIBentryALTinterwordspacing
A.~Adelsbach, S.~Katzenbeisser, and H.~Veith, ``Watermarking schemes provably secure against copy and ambiguity attacks,'' in \emph{{ACM} workshop on Digital rights management}, 2003. [Online]. Available: \url{https://doi.org/10.1145/947380.947395}
\BIBentrySTDinterwordspacing

\bibitem[Behnezhad(2023)]{Behnezhad23}
\BIBentryALTinterwordspacing
S.~Behnezhad, ``Dynamic algorithms for maximum matching size,'' in \emph{{ACM-SIAM} Symposium on Discrete Algorithms, {SODA}}.\hskip 1em plus 0.5em minus 0.4em\relax {SIAM}, 2023. [Online]. Available: \url{https://doi.org/10.1137/1.9781611977554.ch6}
\BIBentrySTDinterwordspacing

\bibitem[Solomon(2016)]{Solomon16}
\BIBentryALTinterwordspacing
S.~Solomon, ``Fully dynamic maximal matching in constant update time,'' in \emph{{IEEE} Annual Symposium on Foundations of Computer Science, {FOCS}}.\hskip 1em plus 0.5em minus 0.4em\relax {IEEE} Computer Society, 2016. [Online]. Available: \url{https://doi.org/10.1109/FOCS.2016.43}
\BIBentrySTDinterwordspacing

\bibitem[Ji et~al.(2021{\natexlab{b}})Ji, Ayday, Yilmaz, and Li]{ji2022differentially}
\BIBentryALTinterwordspacing
T.~Ji, E.~Ayday, E.~Yilmaz, and P.~Li, ``Differentially-private fingerprinting of relational databases,'' \emph{CoRR}, vol. abs/2109.02768, 2021. [Online]. Available: \url{https://arxiv.org/abs/2109.02768}
\BIBentrySTDinterwordspacing

\end{thebibliography}
\end{document}